\documentclass[submission,copyright,creativecommons]{eptcs}
\providecommand{\event}{FOCLASA 2015}

\usepackage{latexsym,amssymb,url,boxedminipage}
\usepackage{amsmath}
\usepackage{amssymb}
 \usepackage{times}
 \usepackage{enumerate}
\usepackage{xspace}
\usepackage{mathpartir}

\newcommand{\chor}[3]{#1_{#2\rightarrow #3}}

\newcommand{\pinull}{\mathbf{0}}

\newcommand{\parop}{\,|\,}      

\makeatletter
\def \rightarrowfill{\m@th\mathord{\smash-}\mkern-6mu%
  \cleaders\hbox{$\mkern-2mu\mathord{\smash-}\mkern-2mu$}\hfill
  \mkern-6mu\mathord\rightarrow}
\makeatother

\newcommand{\bigfract}[2]{\frac{^{\textstyle #1}}{_{\textstyle #2}}}

\newcommand{\trans}[1]{\stackrel{#1}{\longrightarrow}}

\newcommand{\pisucc}{\mathbf{1}}

\newcommand{\pa}{{|\!|}}
\newcommand{\vuota}{\varepsilon}
\newcommand{\sem}[1]{[\![#1]\!]}

\newtheorem{adefinizione}{Definition}[section]
\newtheorem{fatto}[adefinizione]{Fact}
\newtheorem{alemma}[adefinizione]{Lemma}
\newtheorem{teorema}[adefinizione]{Theorem}
\newtheorem{corollario}[adefinizione]{Corollary}
\newtheorem{proposizione}[adefinizione]{Proposition}
\newtheorem{esempio}[adefinizione]{Example}

\newenvironment{definition}{\begin{adefinizione}\ \rm}{\end{adefinizione}}

\newenvironment{theorem}{\begin{teorema}\ \rm}{\end{teorema}}

\newenvironment{example}{\begin{esempio}\ \rm}{\end{esempio}}

\newcommand{\evol}[1]{\ensuremath{\mathcal{E}^{\mathtt{#1}}}\xspace}
\newcommand{\til}[1]{\widetilde{#1}}
\newcommand{\outC}[1]{\overline{#1}}  % output
\newcommand{\component}[2]{#1 [#2]   }  %a[P]
\newcommand{\update}[2]{\widetilde{#1}\{#2\} } % a{P}
\newcommand{\sepr}{ \  \mid \ }
\def\nil{{\boldsymbol 0}} %nil = 0
\newcommand{\fillcon}[2]{\ensuremath{#1\langle\!\langle #2 \rangle\!\rangle}}  % 1 << 2 >>
\newcommand{\arro}[1]{\xrightarrow[]{#1}}
\newcommand{\rulename}[1]{\textsc{(#1)}}
\newcommand{\andalso}{\quad\quad}
\def\sub#1#2{\{\raisebox{.5ex}{\small$#1$}\! / \mbox{\small$#2$}\}} %substitution

\newcommand{\fand}{\wedge}%and
\newcommand{\forr}{\vee}%or
\newcommand{\Lo}{\mathcal{L}} % logica
\newcommand{\F}{\Lo\xspace}%insieme delle formule
\newcommand{\Fres}{\Lo_{r}\xspace}%insieme delle formule logica ristretta

\newcommand{\At}{\mathrm{At}}
\newcommand{\bpred}{\mathrm{p}}%predicati di base

\newcommand{\sep}{\; \big{|} \;}
\newcommand{\fneg}{\neg}%not
\newcommand{\ev}{\ensuremath{\diamondsuit^*}}
\newcommand{\di}{\ensuremath{\diamondsuit}}
\newcommand{\den}[1]{\ensuremath{[\![{#1}]\!]}}%denotazione{formula}
\newcommand{\sat}{\models}%soddisfa
\newcommand{\state}{{s}}

\newcommand{\OG}{\ensuremath{\mathsf{BA}}\xspace} % bounded adaptation
\newcommand{\LG}{\ensuremath{\mathsf{EA}}\xspace} % eventual adaptation
\newcommand{\evold}[1]{\ensuremath{\mathcal{E}_d^{\mathtt{#1}}}\xspace}
\newcommand{\evols}[1]{\ensuremath{\mathcal{E}_s^{\mathtt{#1}}}\xspace}

\title{Towards Dynamic Updates in Service Composition}
\author{Mario Bravetti
\institute{University of Bologna, Italy / INRIA, France}
\email{mario.bravetti@unibo.it}
}
\def\titlerunning{Towards Dynamic Updates in Service Composition}
\def\authorrunning{M. Bravetti}

\begin{document}
\maketitle
\begin{abstract}

We survey our results about verification of adaptable processes.
We present adaptable processes as a way of overcoming the
limitations that process calculi have for describing patterns of dynamic process evolution.
Such patterns rely on direct ways of controlling the behavior and location of running
processes, and so they are at the heart of the adaptation capabilities present in many
modern concurrent systems. Adaptable processes have named scopes and are sensible to
actions of dynamic update at runtime; this allows to express dynamic and static topologies of adaptable processes as
well as different evolvability patterns for concurrent processes. 
We introduce a core calculus of adaptable processes
and consider verification problems for them: first based on specific properties related to error 
occurrence, that we call bounded and eventual adaptation, and then by considering a 
simple yet expressive temporal logic over adaptable processes.
We provide (un)decidability results of such verification problems over adaptable processes 
considering the spectrum of topologies/evolvability patterns introduced.
We then consider distributed adaptability, where a process can update part of a protocol by
performing dynamic distributed updates over a set of protocol participants. 
Dynamic updates in this context are presented as an extension of our work on choreographies 
and behavioural contracts in multiparty interactions. 
We show how update mechanisms considered for adaptable processes can be used to extend 
the theory of choreography and orchestration/contracts, allowing them to be modified 
at run-time by internal (self-adaptation) or external intervention.

\end{abstract}

%*********************************************************************************
%*********************************************************************************
\section{Introduction}
%*********************************************************************************
%*********************************************************************************

We survey our work about verification of adaptable processes \cite{lmcs,isola,beat} presenting theories that we previously introduced in different contexts and connecting them for the first time. 
We start from adaptation 
mechanisms where updates take place on individual processes only and we classify them according to update patterns and adaptable process topologies. Within such classification we present separation results based un(decidability) of verification of adaptation properties/sets of temporal logic formulae.
Then, we consider a more complex setting
where we exploit our foundational study of adaptable processes to devise update mechanisms 
that involve several participants of a distributed protocol, expressed, for example, by a choreography.

In order to introduce adaptation mechanisms and present a spectrum of topologies/evolvability patterns  we 
consider simple extensions to process calculi.
Process calculi aim at describing formally the behavior of concurrent systems. 
A leading motivation in the development of process calculi has been 
properly capturing the \emph{dynamic character} of concurrent behavior.
In fact, much of the success of the $\pi$-calculus~\cite{MilnerPW92a} can be fairly attributed to 
the way it departs from CCS~\cite{Milner89} so as to describe 
mobile systems in which  communication topologies can change dynamically.
Subsequent developments can be explained similarly. 
For instance, the Ambient calculus~\cite{CardelliG00} 
builds on $\pi$-calculus mobility 
to describe the dynamics of interaction within 
boundaries and hierarchies, as
required in distributed systems.
%complements mobility as in the $\pi$-calculus with the 
%dynamics of interaction within 
%with boundaries and hierarchies, thus capturing issues of space/context awareness 
%For instance, 
%the introduction of the Ambient calculus~\cite{CardelliG00} %(and related formalisms) 
%can be justified by the need of describing mobility with considerations of , 
%common in distributed systems.
A commonality in these calculi is that 
the dynamic behavior of a system is realized through a number of \emph{local changes}, 
usually formalized by reduction steps.
Indeed, while in the $\pi$-calculus mobility is enforced by the reconfiguration of individual linkages in the communication topology, % is what characterizes mobility,
in the Ambient calculus spatial mobility is obtained by individual modifications to the containment relations within the ambient hierarchy.
This way, the combined effect of a series of changes at a local level (links, containment relations) 
suffices to explain dynamic behavior at the global (system) level.

%Crucially, in both these formalisms 
%the dynamic behavior of a system is realized through a number of \emph{local changes}, 
%formalized by reduction steps:
%% in its topology:
%while in the $\pi$-calculus the reconfiguration of a single linkage in the communication topology characterizes mobility, 
%in the Ambient calculus  spatial mobility arises from the modification to a single containment relation in the hierarchy of ambients.
%%In both cases, changes at a local level are conceptually lifted to explain concurrent behavior at the global (system) level.
%%This way, the combined effect of local changes suffices to explain dynamic behavior at the global (system) level.
%Consequently, 
%dynamic behavior at the global (system) level can only be obtained 
%from the combined effect of a series of changes at the local level.

There are, however, interesting forms of dynamic behavior that cannot be satisfactorily 
described as a combination of local changes, in the above sense.
These are behavioral patterns which concern change at the \emph{process} level (i.e., the process as a whole), 
and describe \emph{process evolution} along time.
In general, forms of process evolvability are 
characterized by an enhanced control/awareness over 
the current behavior and location of running processes. 
Crucially, this increased control is central to the 
\emph{adaptation} capabilities by which processes modify 
their behavior in response to exceptional 
circumstances in their environment.

This survey reports about the attempt we did to address these shortcomings. In particular, here we present the contents of \cite{lmcs,isola,beat} at a level of detail that allows us to connect them and to discuss the consequences of relating their machinery (this will also lead to new ideas for future work).
In particular, we present, in Section \ref{s:calculi}, a core calculus of adaptable processes
expressing a variety of topologies/evolvability patterns and (un)decidability results of verification problems for them (the technical machinery is taken from \cite{lmcs,isola}).
We will then, in Section \ref{chorupd}, introduce distributed adaptability of protocols as an extension, with update mechanisms similar to those considered for adaptable processes, of the theory of choreography and orchestration/contracts (the technical machinery is taken from \cite{beat}).
Finally, in Section \ref{s:related} we discuss some related work and in Section \ref{s:conclusion} we provide concluding remarks.

%*********************************************************************************
%*********************************************************************************
\section{A Core Calculus of Adaptable Processes: the \evol{} Calculus}\label{s:calculi}
%*********************************************************************************
%*********************************************************************************

%LMCS (BA/EA)+ ISOLA(logica) + BEAT(coreografie con update) + contratti?

We start by presenting \emph{adaptable processes} and results about verification problems for them.
The technical machinery presented in this section is taken from \cite{lmcs,isola} (to which the reader is referred 
for details).

We introduced in \cite{lmcs} the concept of adaptable processes that have a location and 
are sensible to actions of \emph{dynamic update} at runtime.
While locations are useful to designate and structure processes into hierarchies,
dynamic update actions 
%allow to implement forms of process evolvability.
implement a sort of built-in adaptation mechanism.
We illustrate this novel concept by 
 introducing \evol{}, a core process calculus of adaptable processes. % (Section \ref{s:calculi}).
The \evol{} calculus arises as
 a variant of CCS without restriction and relabeling, and 
extended with primitive notions of \emph{location} and \emph{dynamic update}.
In $\mathcal{E}$,  
$\component{a}{P}$ denotes the adaptable process $P$ located at $a$.
Name $a$ acts as a \emph{transparent} locality: $P$ can 
evolve on its own but also  
interact freely with its environment.
Localities can be nested, and are sensible to interactions with \emph{update prefixes}.
An update prefix $\update{a}{U}$ decrees the update of the adaptable process at $a$
with the behavior defined by 
%\emph{update patterns}
$U$, a \emph{context} with zero or more holes,  denoted by $\bullet$.
The \emph{evolution} of $\component{a}{P}$  is realized by 
its interaction  
with the update prefix 
$\update{a}{U}$, which leads to 
%process 
%$U \sub P \bullet$, 
%, which represents 
%$U[P]$, 
%i.e., 
the process obtained by replacing every hole $\bullet$ in $U$ by $P$, denoted 
\fillcon{U}{P}.

%We investigate the decidability of 
%%bounded and eventual adaptation 
%\OG and \LG 
%in 
We consider several variants of \evol{},  obtained via  
two orthogonal characterizations.
%The properties of Evolvable CCS depend on its syntax and on restrictions on such a syntax. 
%\begin{list}{\labelitemii}{\leftmargin=1em}
%\item 
The first one is 
\emph{structural}, and % characterization
%distinguishes between
distinguishes between
\emph{static}  and \emph{dynamic} topologies of adaptable processes. The former is enforced by constraining the latter so that the topology of adaptable processes $\component{a}{P}$ not to vary along the evolution of the system:
they cannot be destroyed nor new ones can appear. %, by e.g. putting them after a prefix.
%Hence, for example, in a static topology, any $\update{a}{U}$ is required to be such that $U = \component{a}{U'}$, thus
%any interaction of $\component{a}{P}$ with $\update{a}{U}$
%leads necessarily to component $\component{a}{U \sub P \bullet}$. 
%In contrast, in the more general dynamic topology this restriction is lifted. 
%We will use the subscripts $s$ and $d$ to denote the 
%variants of \evol{} with static and dynamic topologies,
%respectively.
The second characterization is 
%\item A 
\emph{behavioral}, and  %characterization  
concerns 
%features of  of 
\emph{update patterns}---the 
context 
$U$ in an update prefix $\update{a}{U}$.
As hinted at above, update patterns determine the 
 behavior of running processes after an update action.
In order to account for different evolvability patterns, 
we consider three kinds of update patterns, which determine three 
families of \evol{} calculi---denoted by the superscripts 
1, 2, and 3, respectively.
%\evol{1}, \evol{2}, and \evol{3}, respectively.
%\footnote{The family \evol{2} has been called \evol{-} in abridged versions of the present work (\cite{BGPZFMOODS}).}
%The first update pattern 
%admits all kinds of contexts, and so it represents  
%the most expressive form of update. 
%In particular, holes $\bullet$ can appear behind prefixes.
%The second update pattern forbids such guarded holes in contexts. 
%In the third update pattern we further require contexts to have exactly one hole, 
%thus preserving the current behavior (and possibly adding new behaviors): 
%this is the most restrictive form of update. 
%%\end{list}
%\begin{example}
%\input{example}\qed
%\end{example}
%\todo{motivation for the static/dynamic distinction??}
In our view, 
%the variants of  \evol{}  derived from these two characterizations 
these variants
capture a fairly ample spectrum of scenarios that arise in the 
joint
analysis of correctness and adaptation concerns in evolvable systems.
%component aggregations.
They borrow inspiration from 
existing programming languages,
development frameworks, and 
component models.
We now present the $\evol{}$ calculus, its different variants, and its operational semantics. 
We refer to~\cite{lmcs} for further details and discussions.
The $\evol{}$ calculus %in the sequel) 
is a variant of CCS \cite{Milner89} without restriction and relabeling, 
and extended with constructs for evolvability. 
As in CCS, in $\evol{}$, 
processes can perform actions or synchronize on them.  
We presuppose a countable
set $\mathcal{N}$ of names, ranged over by $a,b, \ldots$, possibly decorated as 
$\overline{a},  \overline{b}$ and $\til{a}, \til{b}$. 
As customary, we use $a$ and $\outC{a}$ to denote atomic input and output actions, respectively.
The syntax of $\evol{}$ processes 
extends that of CCS with 
 primitive notions of \emph{adaptable processes} $\component{a}{P}$
 and \emph{update prefixes} $\update{a}{U}$.
\begin{definition}\label{d:finiteccs}
The classes of $\evol{}$ \emph{processes}, 
\emph{prefixes}, and \emph{update patterns}
are described by the following
grammars: 
%$$
%P        ::= \component{a}{P} \sepr
%          P \parallel P  \sepr ! \pi.P \sepr \sum_{i \in I} \pi_i.P  \qquad \quad  \pi   ::=  a \sepr \outC{a} \sepr \update{a}{U} 
%$$
%%where \emph{update patterns} $U$,
%%::=P_{\bullet}$, 
%%are defined by:
%%as in Definition \ref{d:contexts}.
%$$
%U        ::= \component{a}{U} \sepr U \parallel U  \sepr ! \pi.U \sepr \sum_{i \in I} \pi_i.U \sepr \bullet
%$$
\begin{eqnarray*}
P        &::= & \component{a}{P} \sepr
          P \parallel P  \sepr ! \pi.P \sepr \sum_{i \in I} \pi_i.P  \qquad \quad  \pi   ::=  a \sepr \outC{a} \sepr \update{a}{U} \\
U        & ::= & \component{a}{U} \sepr U \parallel U  \sepr ! \pi.U \sepr \sum_{i \in I} \pi_i.U \sepr \bullet
\end{eqnarray*}

\end{definition}
%Above, the $U$ in the update prefix
%$\update{a}{U}$ is an \emph{update pattern}: it 
Intuitively, update patterns above
represent
a context, i.e., a process with zero or more \emph{holes}.
% (see Definition \ref{d:contexts} below).
The intention is that when an update prefix 
$\update{a}{U}$
is able to interact, 
the current state of  an adaptable process named $a$ 
is used to fill the holes in the update pattern $U$. 
Given a process $P$, 
process $\component{a}{P}$ denotes 
the adaptable process $P$ \emph{located at} $a$.
Notice that $a$ acts as a \emph{transparent} locality: process $P$ can evolve on its own, and  interact freely with external processes.
Localities can be nested, so as to form suitable hierarchies of adaptable processes.
The rest of the syntax follows standard lines.
A process $\pi.P$ performs prefix $\pi$ and then behaves as $P$. 
Parallel composition $P \parallel Q$ decrees the concurrent execution of $P$ and $Q$.
We abbreviate $P_{1} \parallel \cdots \parallel P_{n}$ as $\prod_{i=1}^{n} P _{i}$, and
use $\prod^{k} P$ to denote the parallel composition of $k$ instances of process $P$.
Given an index set $I = \{1,..,n\}$, the guarded sum $\sum_{i \in I} \pi_{i}.P_{i}$ represents an exclusive choice
over $\pi_{1}.P_{1}, \ldots, \pi_{n}.P_{n}$.
As usual, we write $\pi_{1}.P_{1} + \pi_{2}.P_{2}$ if ${|}I{|}=2$, and $\nil$ if $I$ is empty. 
Process $!\, \pi.P$ defines guarded replication, i.e., 
unboundedly many occurrences of $P$ in parallel, which are triggered by prefix $\pi$.
 
%We now define  a general way 
%of extending the grammar of process languages with holes, so as to 
%define update patterns.
%Intuitively, we extend rule productions with a hole (denoted 
%$\bullet$), distinguishing between rule productions
%for process expressions (so-called \emph{process categories}) from 
%the rest.
%In particular, we would like to avoid adding holes to rule 
%productions for prefixes (i.e., productions for $\pi$ in the 
%syntax).

%\begin{definition}\label{d:contexts}
%Given a process category $E$, we denote with $E_{\bullet}$ the 
%process category with rule productions obtained from those of $E$ 
%by: 
%\begin{enumerate}
%\item adding a new rule ``$E_{\bullet} ::= \bullet$''; 
%\item replacing every rule ``$E ::= term$'' of $E$
%with a rule ``$E_{\bullet} ::= term_{\bullet}$'', 
%where ``$term_{\bullet}$'' is obtained from ``$term$'' by 
%syntactically replacing all process categories $F$ occurring in ``
%$term$'' by $F_{\bullet}$. 
%\end{enumerate}
%\end{definition}
 
 Given an update pattern $U$ and a process $Q$, we write $\fillcon{U}{Q}$ for 
 the process obtained by filling in with $Q$ those holes in $U$ not occurring inside update prefixes (a formal definition can be found in~\cite{lmcs}). %Notice that 
% $\fillcon{U}{Q}$ 
 %Indeed, it is worth observing that Definition \ref{d:fillit} 
% does not replace holes inside prefixes, which 
%ensures a consistent treatment of nested update actions.
Hence, $\{ \cdot \}$ can be seen as a scope delimiter for holes $\bullet$  in $\update{a}{U}$.
%Indeed, it is worth observing that Definition \ref{d:fillit} does not replace holes inside prefixes; 
%this ensures a consistent treatment of nested update actions.

%\todo{Jorge: We could remove the following definition, leave intuitions, and point to the TR instead.}
%
%\begin{definition}\label{d:fillit}
%The effect of replacing the holes in 
%an update pattern $U$ with a process $Q$, 
%denoted \fillcon{U}{Q}, 
% is defined inductively on $U$ as follows:
%\begin{align*}
%\fillcon{\bullet\,}{Q} &= Q & 
%\fillcon{(U_{1} \parallel U_{2})}{Q} & =  \fillcon{U_{1}\,}{Q} \parallel \fillcon{U_{2}\,}{Q} \\
%\fillcon{\component{a}{U}}{Q} &= \component{a}{\fillcon{U}{Q}} &
%\fillcon{\bigg(\sum_{i \in I}\pi_{i}.U_{i}\bigg)}{Q} &= \sum_{i \in I}\pi_{i}.\fillcon{U_{i}\,}{Q} \\
%\fillcon{(! \pi.U)}{Q} &= !\pi.(\fillcon{U}{Q})
%\end{align*}
%
%\end{definition}

We now move on to consider 
three concrete instances of 
update patterns $U$. 

\begin{definition}\label{def:varianti}
We shall consider the following three instances of update patterns for \evol{}:
\begin{enumerate}
\item {\bf Full $\evol{}$ (\evol{1}).}
The first 
update pattern
admits all kinds of contexts for update prefixes. %i.e., $U ::= P_{\bullet}$.
This variant, corresponding to the above \evol{} is
 denoted also with \evol{1}.

\item {\bf Unguarded $\evol{}$ (\evol{2}).}
In the second
update pattern, 
holes cannot occur in the scope of prefixes in $U$:
$$
U  ::=  P \sepr \component{a}{U}  \sepr    U \parallel U  \sepr \bullet  %$
$$
The variant of \evol{} that adopts this update pattern is denoted
\evol{2}.
\item {\bf Preserving $\evol{}$ (\evol{3}).}
In the third update pattern, the 
current state of the adaptable process is always preserved (i.e. 
``$\bullet$'' must occur exactly once in $U$).
Hence,  
it is only possible to add new adaptable processes and/or behaviors in parallel or to relocate it:
$$
U  ::=  \component{a}{U}  \sepr U \parallel P \sepr \bullet
$$
The variant of \evol{}  that adopts this update pattern is denoted 
\evol{3}.

\end{enumerate}

\end{definition}

%The semantics of \evol{} processes 
The process semantics
is given in terms of a 
Labeled Transition System (LTS). It 
is  generated by the set of rules in Figure~\ref{fig:ltswithalpha}.
%We  introduce some auxiliary definitions first.
In addition to the standard CCS actions (input, output, $\tau$), we consider two
complementary actions for process update: 
$\update{a}{U}$ and $\component{a}{P}$.
The former represents the possibility to enact an update pattern $U$ for the adaptable process at $a$;
the latter 
says that 
%expresses the fact that 
an adaptable process at $a$, with current state $P$, 
can 
%possibly 
be updated. 
We define $\arro{~~~}$ as $\arro{~\tau~}$, and write $P \arro{~\alpha}$ if $P \arro{~\alpha} P'$, for some $P'$.

%\begin{definition}\label{d:srtcong}
%\emph{Structural congruence} is  the smallest
%congruence relation generated by the following laws: 
%$P \parallel Q \equiv Q \parallel P$; $P \parallel (Q \parallel R) \equiv (P \parallel Q) \parallel R$.
%\end{definition}

%\begin{definition}[Normal Form]\label{d:nform}
%An \evol{} process $P$ is said to be in \emph{normal form} iff
%$$P = \prod_{i=1}^{m} P_i \parallel \prod_{j=1}^{n} \component{a_j}{P'_{j}} $$
%  where, for $i \in \{1,\ldots,m\}$, $P_i$ is not in the form $Q \parallel Q'$ or $a[Q]$, 
%  and, for all $j \in \{1,\ldots,n\}$,  %${P'_{j,k}}$ 
%${P'_{j}}$ is in normal form.
%Note that if $m = 0$ then 
%the normal form is simply 
%$P = \prod_{j=1}^{n} \component{a_j}{P'_{j}}$; similarly, if $n=0$ then the normal form is $P = \prod_{i=1}^{m} P_i$.
%\end{definition}
%
% \begin{lemma}\label{lem:normalform}
% Every \evol{} process
% is structurally congruent
%  to a process in normal form.%, i.e. 
% \end{lemma}
% 
%
%%We are now ready to define an LTS semantics for \evol{}. 
%The LTS 
%for \evol{} processes
%is  generated by the set of rules in Figure \ref{fig:ltswithalpha}.

\begin{definition}\label{d:lts}
The LTS for \evol{}, denoted $\arro{~\alpha~}$,  
is defined by the rules in Figure~\ref{fig:ltswithalpha}, with transition labels defined as:
%Given transition labels 
\[
\alpha    ::=  ~~ a \sepr \outC{a} \sepr \component{a}{P} \sepr \update{a}{U} \sepr \tau
\]
\end{definition}

% \begin{remark}\label{r:fbranch}
% The LTS for \evol{} is finitely branching. 
% The proof proceeds by induction on the syntactic structure of terms; the base cases are $\sum_{i\in I}\pi_i.U_i$ and $!\pi.U$.
% \end{remark}

\begin{figure}[t]
$$
\inferrule[\rulename{Sum}]{}{\sum_{i\in I} \pi_i.P_i \arro{~\pi_j~}  P_j  ~~(j \in I)}  
\qquad
\inferrule[\textsc{(Repl)}]{}{!\pi.P \arro{~\pi~}  P \parallel !\pi.P }
\qquad
\inferrule[\rulename{Comp}]{}{\component{a}{P} \arro{~\component{a}{P}~}  \star}
\qquad 
\inferrule[\rulename{Loc}]{P \arro{~\alpha~} P'}{\component{a}{P} \arro{~\alpha~}  \component{a}{P'}}
$$
$$
\inferrule[\rulename{Act1}]{P_1 \arro{~\alpha~} P_1'}{P_1 \parallel P_2 \arro{~\alpha~} P'_1 \parallel P_2}			
\qquad
\inferrule[\rulename{Tau1}]{P_1 \arro{~a~} P_1' \andalso P_2 \arro{~\outC{a}~} P'_2}{P_1 \parallel P_2 \arro{~\tau~}  P'_1 \parallel P'_2}
\qquad
\inferrule[\rulename{Tau3}]{P_1 \arro{~\component{a}{Q}~} P_1'\andalso P_2 \arro{~\update{a}{U}~} P_2'  }
{P_1 \parallel P_2 \arro{~\tau~} P_1'\sub{ \fillcon{U}{Q}  }{\star} \parallel P_2'}
$$

\caption{LTS for \evol{} .
Rules \rulename{Act2}, \rulename{Tau2}, and \rulename{Tau4}---the symmetric counterparts of 
\rulename{Act1}, \rulename{Tau1}, and \rulename{Tau3}---have been omitted.} \label{fig:ltswithalpha}
\end{figure}

In Figure \ref{fig:ltswithalpha}, rules \rulename{Sum}, \rulename{Repl}, 
\rulename{Act1},
and \rulename{Tau1} 
 are standard.
Rule 
\rulename{Comp} represents the contribution of a process at $a$ in an update operation; 
we use  $\star$ to denote a unique placeholder.
Rule \rulename{Loc}
formalizes
transparency of localities.
Rule \rulename{Tau3} formalizes process evolvability.
To realize the 
evolution of an adaptable process at $a$, it requires: 
(i)  a process $Q$---which represents its current state; 
(ii) an update action offering an update pattern  $U$ for updating the process at $a$---which is represented in $P'_{1}$ by $\star$ (cf. rule \rulename{Comp})
%;
%(iii) that $\mathsf{cond}(U,Q)$ holds (cf. Definition \ref{d:lts}).
As a result,  
$\star$ in $P'_{1}$ is replaced with process $\fillcon{U}{Q}$. Notice that this means that 
%the adaptable process 
the locality 
being updated is discarded unless it is re-created by $\fillcon{U}{Q}$. 
% (cf. Definition \ref{d:fillit}).

% The following lemma is standard:
% \begin{lemma}
% Let $P$ be an \evol{} process.
% Structural congruence is preserved by reduction: 
% if $P \equiv Q$ and $P \pired P'$, then also $Q \pired Q'$ for some $P' \equiv Q'$.
% \end{lemma}

%\todo{Jorge: The following paragraph is included identically in Sect 5 (Def 14, Conv 1)}
We introduce some definitions that will be useful in the following.
%$\rightarrow^{+}$ (resp.
We denote with $\rightarrow^*$ the
%transitive (resp. the
reflexive and transitive
%)
closure of the relation $\rightarrow$.
We define $Pred(s)$ as the set $\{s' \in S \mid s' \rightarrow s \}$ of \emph{immediate predecessors} of $s$,
while $Pred^*(s)$ 
%and $Pred^+(s)$ 
denotes the set $\{s \in S \mid s' \rightarrow^* s\}$ 
%and  $\{s \in S \mid s' \rightarrow^+ s\}$, respectively, 
of \emph{predecessors} of $s$. We will also assume point-wise extensions of such definitions to sets, i.e.
$Pred(S) = \bigcup_{s \in S} Pred(s)$ and similarly for $Pred^*(S)$. 

%*********************************************************************************
\subsection{The static \evol{} Calculus}\label{s:static}
%*********************************************************************************
\emph{Static} variants of \evol{} are defined in~\cite{lmcs}. 
Informally speaking, the static characterization of \evol{} processes is related to tree-like structures obtained
by nesting of located processes. 
In \emph{dynamic} adaptable processes, 
the class here considered, 
update actions which modify the nesting structure are allowed; 
in contrast, in static adaptable processes such actions are disallowed: 
this guarantees that no adaptable process is created nor destroyed along computation. 
By adding such a static topology constraint we get three static variants \evols{1}, \evols{2} and \evols{3} for the three kind of update patterns considered in the \evol{1}, \evol{2} and \evol{3} languages. From now on we will denote the latter with \evold{1}, \evold{2} and \evold{3} to distinguish them from their static variants.

%*********************************************************************************
\subsection{Eventual and Bounded Adaptation}\label{s:properties}
%*********************************************************************************

In~\cite{lmcs}, we study  two  \emph{verification problems} associated to $\mathcal{E}$ processes and their (un)decidability.
%We propose two such problems. 
%correctness properties for $\mathcal{E}$ processes.
They are defined in terms of standard observability predicates (\emph{barbs}), which indicate the presence of a designated error signal.
We thus distinguish between \emph{correct states} (i.e., states in which no error barbs are observable) and \emph{error states} (i.e., states exhibiting error barbs).
The first verification problem, 
\emph{bounded adaptation} 
(abbreviated \OG)
ensures that,
given a finite $k$, 
at most $k$ consecutive error states can arise  in computations of the system---including those reachable as a result of
dynamic updates. In other words,
the number of consecutive erroneous states that can be traversed
during a computation is bound by some given number $k$.
The second one, 
\emph{eventual adaptation} (abbreviated \LG), 
is similar but weaker: 
it ensures that if the system enters into an error state then it will eventually reach a correct 
state (the system cannot enter a sequence of consecutive erroneous states that lasts forever). %---that is,  it ensures that only finitely many errors can occur.
We believe that 
\OG and \LG
%bounded and eventual adaptation 
fit well in 
the kind of correctness analysis that is required in a number of emerging applications.  
%including, but not limiting to, cloud computing scenarios.
%Returning to the cloud computing setting motivated at the beginning, 
%these properties allow to check, 
For instance, 
on the provider side of a cloud computing application, %it is possible to check 
these properties allow to check
whether a 
client is able to assemble faulty systems
via the aggregation of the provided services and the possible subsequent updates.
On the client side, it is possible to 
%perform forms of 
carry out forms of 
\emph{traceability analysis}, so as to 
prove that if 
%an incorrect computation has been executed by the system, 
the system exhibits an incorrect behavior, then 
it follows from a bug in the provider's infrastructure
and not from the initial aggregation and dynamic updates 
provided by the client.

%In addition to  
%error occurrence, 
%%to characterize forms of error, 
%the correctness of
%adaptable processes 
%must consider the fact 
%%it is fundamental to take into account 
%that the number of 
%modifications (i.e. update actions)
%that can be 
%applied to the system is typically \emph{unknown}. 
%For this reason, we 
%consider \OG and \LG in conjunction with 
%the notion of \emph{cluster} of adaptable processes.
%Given a system $P$ and a set $M$ of possible updates that
%can be applied to it at runtime, 
%its associated cluster considers $P$ together 
%with an arbitrary number of instances of the updates in $M$.
%%that can be performed over $P$ at runtime.
%This way, 
%a cluster  formalizes adaptation and correctness properties 
%of an initial system configuration (represented by an aggregation of adaptable processes)
%in the presence of arbitrarily many sources of update actions.
%For instance, in a cloud computing scenario %sketched earlier:
%the notion of cluster captures 
%the cloud application as initially deployed by the client along with 
%the options offered by the provider for 
%its evolution at runtime. 
%
%

The main technical results of the paper are summarized in Table \ref{t:results}.
The calculus \evol{1} is shown to be Turing complete, 
%thus evidencing the expressive power of update actions.
%As a result, 
and both \OG and \LG are shown to be 
\emph{undecidable} for \evol{1} processes. % (Section~\ref{s:ev1}).
The Turing completeness of \evol{1} says much on the expressive power of update actions.
In fact, it is known that fragments of CCS without restriction
can be translated into finite Petri nets, %(see, e.g.,  the discussion in \cite{Bravetti09})
so they are not Turing complete.
Update actions in \evol{} thus allow to ``jump''  from finite Petri nets to a Turing complete model.
We show that in \evol{2} 
\OG is \emph{decidable}, % (Section~\ref{sec:decwsts}), 
while \LG remains \emph{undecidable}. % (Section~\ref{sec:unde2d}).
Interestingly,  \LG is already undecidable in \evold{3}, while it is \emph{decidable} in \evols{3}.

\begin{table}[t]
%\begin{figure}[t]
%\begin{floatingtable}[r]{
\begin{center}
\begin{tabular}{c|c|c}
		& \evold{} -- Dynamic Topology & \evols{} -- Static Topology \\
\hline \hline
\evol{1}	&~ \OG undec ~/~\LG undec ~& ~\OG undec~/~\LG undec~\\
\hline
\evol{2}	& ~ \OG dec~/~\LG undec ~ & ~ \OG dec~/~\LG undec \\
\hline
\evol{3}	& ~ \OG dec~/~\LG undec ~ & \OG~dec~/~\LG dec 
\end{tabular}
\end{center}
\caption{\label{t:results} Summary of  (un)decidability results for dialects of \evol{}.}
%We use \evol{i} to denote the family of \evol{} with the $i$-th update pattern.
%\end{figure}
%\end{floatingtable}
% \linefigure
\end{table}

%*********************************************************************************
\subsection{A Logic for Adaptable Processes}\label{s:logic}
%*********************************************************************************

%\input{logic}
\newcommand{\evv}{\ensuremath{\di^{+}}}
\newcommand{\defi}{\ensuremath{\stackrel{\mathsf{def}}{=}}}

%\subsection{Definitions}

We also considered, in \cite{isola}, a 
simple yet expressive temporal logic $\F$ over adaptable processes. Concerning model-checking with formulas of the entire logic $\F$ we have the same decidability results as for the eventual adaptation property, while for a fragment of the logic, denoted with $\Fres$, we have the same decidability results as for the bounded adaptation property (see Table \ref{t:results}).

\begin{definition}
The set $\At$ of atomic predicates $\bpred$ is given by the following syntax:\;
\begin{center}
$\bpred\,::=\,a\sep\outC{a}\sep T$.
\end{center}
\end{definition}

Predicates $a$ and $\outC{a}$ hold true for states/terms that may perform transitions $a$ and $\outC{a}$, respectively. 
The intention is that the interpretation of atomic predicates should coincide with the notion of \emph{barb} in the process model.
$T$ is the true predicate that holds true for every state/term.
In the following, we use $\alpha$ to range over labels $a, \outC{a}$, for some name $a$. % also called \emph{barbs}. 

%we consider the same logic as in the Acciai/Boreale paper without
%the spacial operator (i.e. a very simple logic with a diamond star 
%operator). we consider an extended version of the logic where we 
%have also a simple diamond operator looking at the next step
\begin{definition}\label{def:logic}
The set $\F$ of  logic formulae $\phi,\psi,\ldots$ is given by the following syntax, where $\bpred\in \At$:\;
\begin{center}
%{\small
$\phi\,::=\,\bpred\sep \phi\forr\phi \sep \phi\fand\phi \sep\fneg\phi \sep \di\phi \sep \ev\phi $
%}
\end{center}
\end{definition}

The set of logical operators includes 
%spatial modalities
atomic predicates $\bpred\in\At$,
the usual boolean
connectives ($\forr$, $\fand$, and $\neg$), as well as 
dynamic connectives (the next and eventuality modalities, $\di$ and $\ev$). 
The interpretation of 
$\F$ over LTSs is given below, where each formula is mapped into the set of states/terms satisfying it.
%
%\begin{table}[!t]

{\small
\begin{center}
$\begin{array}{r@{\:}c@{\:}l@{\qquad}r@{\:}c@{\:}l@{\qquad}r@{\:}c@{\:}l@{\qquad}r@{\:}c@{\:}l}
\den{\alpha}&=&  \big\{\state\in \evol{}\,|\, \state \arro{~\alpha~}
\big\}& 
\den{T}&=& \evol{}& 
\den{\di\phi}&=&Pred(\den{\phi})&
\den{\ev\phi}&=&Pred^{*}(\den{\phi})
\vspace*{0.1cm}\\
\multicolumn{12}{c}{
\den{\phi_1\forr\phi_2}\:=\: \den{\phi_1}\cup\den{\phi_2}\qquad
\den{\phi_1\fand\phi_2}\:=\:\den{\phi_1}\cap\den{\phi_2}\qquad
\den{\neg\phi}\:=\:\evol{}\setminus\den\phi
}\vspace*{0.1cm}\\
%\multicolumn{9}{c}{\den{\phi_1|\phi_2}\:=\:\big\{s_1\sumop s_2\,|\,  %\; \state_j\in\den{\phi_j}\text{ for } j=1,2\big\}}
\end{array}$
\end{center}
}
%\caption{Interpretation of formulae over \swsts.}\label{tab:interpretation}
%\end{table}
%
Connectives are interpreted as usual. 
We usually write $\state\sat\phi$  if $\state\in\den\phi$.

\begin{definition}
A formula $\phi$ is called \emph{monotone} if 
%$\phi$ 
it does not contain occurrences of $\neg$.
\end{definition}

Restricted monotone formulae are 
those monotone formulae 
in which conjunctions are always of the form 
$\bpred \fand \phi$, for some $\bpred\in\At$ and a monotone formula $\phi \in \F$.

%additionally constrained to make a restricted use of $\fand$: only $\fand$ with a barb $\alpha$ (or a true predicate) is admitted.
\begin{definition}
A formula $\phi$ is \emph{restricted monotone} if it is monotone and, for any occurrence of $\phi_1\fand\phi_2$ inside $\phi$, there exists $i \in \{1,2\}$ such that $\phi_i$ is a predicate $\bpred\in\At$.
\end{definition}

Restricted logic $\Fres$ is the logic composed by, possibly negated, restricted monotone formulae. 
\begin{definition}
The restricted logic is composed by the set $\Fres$ of  formulae of the form $\phi$ or $\neg \phi$, where
$\phi$ is a \emph{restricted monotone} formula.
\end{definition}

%\todo{
%we observe that some "interesting" properties are expressible in the 
%logic that we consider: i.e. k-boundedness both in the consecutive 
%version, exploiting the simple diamond, (the final one published in the 
%journal) and in the non-consecutive repeated version (but not eventual 
%adaptation and repeated adaptation that would need a recursion operator 
%in the logic)
%}
%
%k-boundedness for $\alpha$ $barb$  (consecutive - forte paper - version)
%
%% Mario version: 
%$\neg \ev \;\; \alpha \fand \ev \;\; \alpha \fand \ev \;\; \alpha \fand \di \;\; \dots \;\; \alpha$, where $\alpha \fand \ev$ is repeated $k$ times
%%$\neg \ev \, \alpha \fand ( \ev \, \alpha \fand (\ev \, \alpha \fand (\ev \, \dots \, \alpha)))$, where $\alpha \fand \ev$ is repeated $k$ times
%
%k-boundedness for $\alpha$ $barb$ (non-consecutive - old - version)
%
%$\neg \ev \;\; \alpha \fand \di \ev \;\; \alpha \fand \di \ev \;\; \alpha \fand \di \ev \;\; \dots \;\; \alpha$, where $\alpha \fand \di \ev$ is repeated $k$ times

%\subsection{Some Example Properties}
We now give some examples of formulas in 
%the expressiveness of 
$\F$ and $\Fres$.
%by means of some examples.
%Below, w
We take $\evv \phi \defi \di \ev \phi$. 
Possibly the most natural safety property one would like to ensure 
is the absence of $k$ \emph{consecutive barbs} (representing, e.g., errors, as for the bounded adaptation property): 
$$\mathsf{CB}_{k}(e) \defi \neg  \ev \big(\underbrace{e \fand  \di(e \fand \di(e \fand \ldots \fand \di e))}_{\text{$e$ appears $k$ times}} \big)$$
Observe how $\mathsf{CB}_{k}(e) \in \Fres$. 
Another example of a sensible property to ensure is \emph{monotone correctness}: once solved, errors do not reappear. In $\F$ this can be expressed as:
$$
\mathsf{MC}(e) \defi \neg \ev \big( e \fand \evv(\neg e \fand \evv e) \big)
$$
Assuming a designated barb $ok$, signaling a correct (error-less) state, 
the above can be captured in $\Fres$ as follows:
$$
\mathsf{MC}^r(ok, e) \defi \neg \ev \big(e \fand \evv(ok \fand \ev e)\big)
$$
The extension of $\mathsf{MC}^r(ok, e)$ to consider $k$ different error 
phases (it cannot happen that an error re-appears up to $k$ times) is 
straightforward:
$$
\mathsf{MC}^r_k(ok, e) \defi \neg \ev \big( 
\underbrace{e \fand \evv(ok \fand \ev (e \fand \evv(ok \fand
\ldots
(ok \fand \ev e))))}_{\text{$ok$ appears $k$ times}} \big)
$$
The latter is used in \cite{isola} to model-check a case study about effective scaling in cloud computing, modeled with a $\evold{2}$ process (thus model-checking is decidable).

In \cite{isola} we also conjecture that decidability results for $\Fres$ can be extended to monotone formulae
(dropping the restricted monotone constraint) and we discuss extensions of the logic with a recursion operator.
In particular, we claim that for $\Fres$ with recursion, that would allow us to express properties like eventual adaptation,
we have the same (un)decidability results as for formulae of $\F$ (without recursion).

%*********************************************************************************
%*********************************************************************************
\section{Choreographies and Orchestrations with Dynamic Updates}\label{chorupd}
%*********************************************************************************
%*********************************************************************************

We now discuss a possible approach for expressing distributed dynamic updates of protocol specifications.
The technical machinery presented in this section is taken from \cite{SC07,beat} (to which the reader is referred 
for details).

Dynamic updates in this context are presented as an extension of our work on choreographies 
and behavioural contracts in multiparty interactions~\cite{fsen07,SC07,fundaInfo08}. 
We consider distributed adaptability, where a process can update part of a protocol by
performing dynamic distributed updates over a set of protocol participants (denoted by {\it roles} in the choreographical description of the protocol).
In particular, we show how update mechanisms considered for adaptable processes can be used to extend 
the theory of choreography and orchestration/contracts, allowing them to be modified 
at run-time by internal (self-adaptation) or external intervention. This entails both an extension of choreography
and orchestration/contract languages with primitives for distributed updates and named adaptable parts similar to operators
$\update{a}{U}$ and $\component{a}{P}$  considered in the previous section. Moreover also related theories must be extended, such as choreography well-formedness (by, e.g., connectedness constraints) and derivation 
of participant orchestrations from a choreography by projection.

We start by presenting the choreography and orchestration calculi (representing individual participant behaviour)
and then we relate them by projection. Finally, we will extend the the theory presented with  distributed dynamic update mechanisms starting from the operators introduced in the previous section.

%*********************************************************************************
\subsection{The Choreography Calculus}
%*********************************************************************************

%Let \emph{Operations}, 
%ranged over by $a, b, c, \cdots$ and \emph{Roles},
%ranged over by $r, s, t, \cdots$, be
%two countable sets of operation and role names, respectively.

We assume a denumerable set of action names ${\cal N}$, ranged over by  $a,b,c,\dots$ and 
a set of roles $Roles$ ranged over by $r, s, l$.

\begin{definition}{\bf (Choreographies)}
The set of $Choreographies$, ranged over by $H, L, \cdots$
is defined by the following grammar:  
\[
H \ \ ::= \ \ \quad a_{r \rightarrow s} \quad | \quad
                    H+H \quad | \quad
                    H;H \quad | \quad
                    H|H \quad | \quad
                    H^*  
\]
The invocations $a_{r \rightarrow s}$ means 
that role $r$ invokes the operation $a$ provided
by the role $s$. The other operators are 
choice $\_+\_$, sequential $\_;\_$, parallel $\_|\_$, 
and repetition $\_^*$.
\end{definition}

%The choreography calculus is used to define
%the possible conversations within a group of
%collaborating services that embody some
%specific roles. The atomic action $a_{r \rightarrow s}$, 
%indeed, represents the invocation from the
%role $r$ of the operation $a$ provided by
%the role $s$.

The operational semantics of choreographies
considers two auxiliary terms
$\pisucc$ and $\pinull$. They are used to model
the completion of a choreography, which is relevant
in the operational modeling of sequential composition.
The formal definition is given in Table~\ref{choreosem} 
where we take $\eta$ to range over the set of 
labels $\{ a_{r \rightarrow s}\ |\ a \in {\cal N}, r,s \in Roles \} \cup \{ \surd\}$
(the label $\surd$ denotes successful completion). 
The rules in Table~\ref{choreosem} are rather standard 
for process calculi with sequential composition
and without synchronization; in fact,
parallel composition simply allows for the interleaving
of the actions executed by the operands (apart from completion labels $\surd$
that are required to synchronize).

\begin{table}[th]
\[
\begin{array}{ccc}
a_{r \rightarrow s} \trans{a_{r \rightarrow s}} \pisucc 
&
\pisucc \trans{\surd} \pinull 
&
H^* \trans{\surd} \pinull 
\\
\\
\bigfract{H \trans{\eta} H'}
         {H\!+\!L \trans{\eta} H'} 
&         
\bigfract{H \trans{\eta} H' \quad \eta \neq \surd} 
         {H;\!L \trans{\eta} H';\!L}                  
\qquad
&
\bigfract{H \trans{\surd} H' \quad L \trans{\eta} L'}
         {H;\!L \trans{\eta} L'}        
\\
\\
\bigfract{H \trans{\surd} H' \quad L \trans{\surd} L'}
         {H \!\parop\! L \trans{\surd} H' \!\parop\! L'}  
\qquad
&
\bigfract{H \trans{\eta} H' \quad \eta \neq \surd}
         {H \!\parop\! L \trans{\eta} H' \!\parop\! L}                                     
&
\bigfract{H \trans{\eta} H' \quad \eta \neq \surd}
         {H^* \trans{\eta} H' ; H^*}  
\end{array}
\]

\caption{Semantic rules for choreographies (symmetric rules omitted).}
\label{choreosem}
\end{table}

Choreographies are especially useful to describe the protocols
of interactions within a group of collaborating services.
To clarify this point, we present a simple example of a protocol 
described with our choreography calculus.

%\begin{example}\label{traAgency}
%{\bf (Reservation via Travel Agency)}
%Let us consider the following choreography composed of four
%roles: {\em Client}, {\em TravelAgency}, {\em AirCompany}
%and {\em Hotel}
%\[
%\begin{array}{l}
%Reservation_{Client \rightarrow TravelAgency};\\
%(\ (Reserve_{TravelAgency \rightarrow AirCompany};ConfirmFlight_{AirCompany \rightarrow TravelAgency})\ | \\
%\ \
% (Reserve_{TravelAgency \rightarrow Hotel};ConfirmRoom_{Hotel \rightarrow TravelAgency})\ );\\
%Confirmation_{TravelAgency \rightarrow Client}+
%Cancellation_{TravelAgency \rightarrow Client} 
%\end{array}
%\]
%\end{example}

\begin{example}\label{BSB}
{\bf (Buyer/Seller/Bank)}
Let us consider the following choreography composed of three
roles: {\em Buyer}, {\em Seller} and {\em Bank}
\[
\begin{array}{l}
\mathit{Request}_{\mathit{Buyer} \rightarrow \mathit{Seller}};
(\ \mathit{Offer}_{\mathit{Seller} \rightarrow \mathit{Buyer}} \parop
\mathit{PayDescr}_{\mathit{Seller} \rightarrow \mathit{Bank}}\ ); \\
\hspace{.5cm} Payment_{\mathit{Buyer} \rightarrow \mathit{Bank}};
(\ \mathit{Confirm}_{\mathit{Bank} \rightarrow \mathit{Seller}} \parop 
\mathit{Receipt}_{\mathit{Bank} \rightarrow \mathit{Buyer}}\ )
\end{array}
\]
According to this choreography, the {\em Buyer} initially
sends an item request to the {\em Seller} that subsequently,
in parallel, replies with an offer and sends a payment description to the {\em Bank}.
Then the {\em Buyer} performs the actual payment to the {\em Bank}. The latter in parallel
replies with a receipt and sends a payment confirmation to the {\em Seller}.
\end{example}

%*********************************************************************************
\subsection{The Orchestration Calculus}
%*********************************************************************************

%The set ${\cal N}_{con} = \{ a_{*} \; | \; a \in {\cal N} \}$ is the set of contract action names.
%We take $\alpha$ to range over the set of all names
%${\cal N}_{all}$ = 
%${\cal N} \cup {\cal N}_{con}$.
%Moreover, we consider a denumerable set ${Roles}$ of role names,
%ranged over by $l, l', l_1, \cdots$. 
%, which is disjoint from the set of names ${\cal N}$. 
%The set ${\cal N}_{loc} = \{ a_l \; | \; a \in {\cal N}, l \in {Roles} \}$ is the set of located action names.
%We take $\alpha$ to range over the set of all names
%The set ${\cal N}_{out} = {\cal N}_{con} \cup {\cal N}$ is the set of output names.
%\{ a_{*} \; | \; a \in {\cal N} \}$ is the set of internal names.
%The set ${\cal A} = \{ a_l, \overline{a_l} \; | \; a \in {\cal N}, l \in {Roles} \}$
%is the set of input and output actions. 
%The set ${\cal A}_{con} = {\cal N}_{con} \cup \{ \overline{a}_* \; | \; a_* \in {\cal N}_{con} \}$
%is the set of input and output contract actions. 
%The set ${\cal A}_{loc} = {\cal N}_{loc} \cup \{ \overline{a}_l \; | \; a_l \in {\cal N}_{loc} \}$
%is the set of input and output located actions. 
%The set ${\cal A}_{all} = {\cal N}_{all} \cup \{ \overline{\alpha} \; | \; \alpha \in {\cal N}_{all} \}$
%is the set of input and output actions. 
As for choreographies, we assume a denumerable set of action names ${\cal N}$, ranged over by  $a,b,c,\dots$. We use $\tau \notin {\cal N}$ to denote an internal (unsynchronizable) computation. 

%SERVE QUESTO?
%
%
%Given a set of located action names $I \subset {\cal N}_{loc}$, 
%we denote: with $\overline{I} = \{ \overline{a}_l \ |\ a_l \in I\}$ 
%the set of output actions performable on those names and
%with $I_l = \{ a\ |\ a_l \in I\}$ the set of action names with
%associated role $l$.
%Given a set of (input and output) located actions $L \subset {\cal A}_{loc}$,
%we denote with $L_l$ the set of 

\begin{definition}\label{orcsys} {\bf (Orchestrations and Systems)}
The syntax of orchestrations is defined by the following
grammar 
\[
\begin{array}{lll}
C &\ ::=\ &      \pinull \quad | \quad 
               \pisucc \quad | \quad
               \tau  \quad | \quad
%               a_*  \quad | \quad
%               \tau ; \overline{a}_* \quad | \quad
               a  \quad | \quad
%               \tau ; 
		  \overline{a}_r \quad | \quad               
\\[.2cm]
  & &             
               C ; C \quad | \quad
               C\!+\!C \quad | \quad
               C \!\parop\! C \quad | \quad
%               C \!\resop\! M \quad | \quad
               C^* 
\end{array}
\]
%where $M \subseteq {\cal N}_{con}$.
The set of all orchestrations $C$ is denoted by ${\cal P}_{orc}$. Differently from choreographies, orchestration syntax makes it possible to specify orchestrations that fail $\pinull$ or successfully terminate $\pisucc$. The reason of such a difference is that, conceptually, choreographies represent the desired overall system behaviour, while orchestrations can be used to reason about  concrete (possibly failing) participant implementations.
%In the following we will omit trailing ``$\pisucc$'' when writing contracts.
\\
The syntax of systems (parallel composition of orchestrations) is defined by the following
grammar 
\[
P \ \ ::= \ \   [C]_r \quad | \quad
               	P \pa P \quad 
%| \quad
%               	P \re L \quad
\]
%where $L \subseteq {\cal A}_{loc}$. 
A system $P$ is well-formed if: $(i)$ every orchestration subterm 
$[C]_r$ occurs in $P$ at a different role $r$ and $(ii)$ no output action 
with destination $r$ is syntactically included inside an orchestration subterm
occurring in $P$ at the same role $r$, i.e. actions $\overline{a}_r$ cannot occur inside a subterm $[C]_r$
of $P$.
The set of all well-formed systems $P$ is denoted by ${\cal P}$.
In the following we will just consider well-formed systems and, for simplicity, we will call them
just systems. 
%Moreover, we will omit parenthesis``$[ \,]$'' when writing systems.
\end{definition}
We take $\alpha$ to range over
the set of syntactical actions $SAct = {\cal N} \cup \{ \overline{a}_r \; | \; a \in {\cal N} \wedge r \in {Roles} \} \cup \{ \tau \}$.

The operational semantics of orchestrations is defined by the rules
in Table~\ref{contsem} (plus symmetric rules). The operational semantics of systems is defined by the rules
in Table~\ref{contcompsem} plus symmetric rules.
%We take $\gamma$ to range over ${\cal A} \cup \{ \tau \}$ and 
We take $\beta$ to range over
the set $Act$ of actions executable by orchestrations and systems, i.e.
$Act =  SAct \cup
\{ a_l \; | \; a \in {\cal N} \wedge l \in {Roles} \} \cup
\{ a_{r \rightarrow s}\ |\ a \in {\cal N} \wedge r,s \in {Roles} \} \cup 
\{ \overline{a}_{r s}\ |\ a \in {\cal N} \wedge r,s \in {Roles} \}$.
We take $\lambda$ to range over the set of 
transition labels $%{\cal L} = 
Act \cup \{ \surd \}$, where $\surd$
denotes successful termination. 
\begin{table}[tbh]
\[
\begin{array}{cc}
\pisucc \trans{\surd} \pinull 
&
\alpha \trans{\alpha} \pisucc 
%C \oplus D \trans{\tau} C
\\
\\
\multicolumn{2}{c}{
\bigfract{C \trans{\lambda} C'}
         {C\!+\!D \trans{\lambda} C'} 
\qquad         
\bigfract{C \trans{\lambda} C' \quad \lambda \neq \surd} 
         {C;\!D \trans{\lambda} C';\!D}                  
\qquad

\bigfract{C \trans{\surd} C' \quad D \trans{\lambda} D'}
         {C;\!D \trans{\lambda} D'}        
}\\
\\
%\bigfract{C \trans{a_{*}} C' \quad D \trans{\overline{a}_{*}} D'}
%         {C \!\parop\! D \trans{\tau} C' \!\parop\! D'}     
\qquad
\bigfract{C \trans{\surd} C' \quad D \trans{\surd} D'}
         {C \!\parop\! D \trans{\surd} C' \!\parop\! D'}  
\qquad
&
\bigfract{C \trans{\lambda} C' \quad \lambda \neq \surd}
         {C \!\parop\! D \trans{\lambda} C' \!\parop\! D}                           
\\
\\
%\bigfract{C \trans{\lambda} C' \quad \lambda \not\in M \cup \overline{M}}
%         {C \!\resop\! M \trans{\lambda} C' \!\resop\! M}               
C^* \trans{\surd} \pinull 
&
\bigfract{C \trans{\lambda} C' \quad \lambda \neq \surd }
         {C^* \trans{\lambda} C' ; C^*}  
\end{array}
\]

\caption{Semantic rules for orchestrations (symmetric rules omitted).}
\label{contsem}

\end{table}

\begin{table}[tbh]
\[
\begin{array}{clclc}
\bigfract{C \trans{a} C' \quad }
         {[C]_r \trans{a_r} [C']_r} 
& \qquad &
\bigfract{C \trans{\overline{a}_{s}} C' \quad }
         {[C]_r \trans{\overline{a}_{rs}} [C']_r} 
& \qquad &
\bigfract{P \trans{\lambda} P' \quad \lambda \neq \surd }
         {P  \pa Q \trans{\lambda} P' \pa Q} 
\\
\\
\multicolumn{5}{c}{
\bigfract{P \trans{a_s} P' \quad
          Q \trans{\overline a_{rs}} Q'}
         {P \pa Q \trans{a_{r\rightarrow s}} P' \pa Q'} 
\qquad
\bigfract{P \trans{\surd} P' \quad
Q \trans{\surd} Q'}
{P \pa Q \trans{\surd} P' \pa Q'}
}
%\qquad &
%\bigfract{P \trans{\lambda} P' \quad 
%          \lambda \not\in L }
%         {P \re L \trans{\lambda} P' \re L} 
%& &
%\bigfract{P \trans{\lambda} P' \quad 
%          \lambda \in L \cup \overline{L}}
%         {P / L \trans{\tau} P' / L} 
%& &
%\bigfract{P \trans{\lambda} P' \quad 
%          \lambda \notin L \cup \overline{L}}
%         {P / L \trans{\lambda} P' / L}    
\\
\end{array}
\]

\caption{Semantic rules for systems (symmetric rules omitted).}
\label{contcompsem}

\end{table}

Here and in the remainder of the paper we use the following
notations: 
$P \trans{\lambda}$ to mean that there exists $P'$ 
such that $P \trans{\lambda} P'$ and,
given a sequence of labels $w=\lambda_1 \lambda_2 \cdots \lambda_{n-1}\lambda_n$
(possibly empty, i.e., $w = \vuota$), 
we use $P \trans{w} P'$ to denote the sequence of
transitions $P \trans{\lambda_1} P_1 \trans{\lambda_2}
\cdots \trans{\lambda_{n-1}} P_{n-1} \trans{\lambda_n} P'$
(in case of $w=\vuota$ we have $P'=P$, i.e., $P \trans{\vuota} P$).
Moreover, for completely specified systems $P$ (i.e. terms $P$ not included as subterms in a larger term $P'$), we do not consider transitions corresponding to unmatched input and output actions: namely,
we consider only transitions labeled with $\tau$ (local internal
actions), $\surd$ (global successful termination) and 
$a_{r\rightarrow s}$ (completed interactions).

We now define the notion of correct
composition of orchestrations. This notion is the same as in~\cite{fsen07}. Intuitively, a system
composed of orchestrations is correct if all possible
computations may guarantee completion; this 
means that the system is both deadlock and livelock free
(there can be an infinite computation, but given
any possible prefix of this infinite computation, it must be possible to extend it 
to reach a successfully completed computation).

\begin{definition}\label{correctcomposition}
{\bf (Correct orchestration composition)}
System $P \in {\cal P}$ is a correct orchestration composition, 
denoted $P\!\downarrow$,
if for every
$P'$ such that
%\[
$
P \; \trans{w} \; P' 
%\]
$
there exists $P''$ such that 
%\[
$
P' \; \trans{w'} \; P'' \; \trans{\surd}. 
$
%\]
%we have that $P' \trans{\surd}$
\end{definition}

%*********************************************************************************
\subsection{Choreography Implementation, Projection and Well-Formedness}\label{proj}
%*********************************************************************************

We are now ready to formalize the notion of correct implementation 
of a choreography.
With $P \trans{\tau^*} P'$ we denote the existence
of a (possibly empty) sequence of $\tau$-labeled transitions 
starting from the system $P$ and leading to $P'$.
Given the sequence of labels $w = \lambda_1 \cdots \lambda_n$,
we write $P \stackrel{w}{\Longrightarrow} P'$ if 
there exist $P_1, \cdots, P_m$ such that 
$P \trans{\tau^*} P_1 \trans{\lambda_1} P_2 \trans{\tau^*} \cdots \trans{\tau^*}
 P_{m-1} \trans{\lambda_n} P_m \trans{\tau^*} P'$.

Intuitively, a system implements a choreography if it 
is a correct composition of orchestrations and 
all of its conversations  (i.e. 
the possible sequences of message exchanges), 
are admitted by the choreography.

\begin{definition}\label{impl}{\bf (Choreography implementation)}
Given the choreography $H$ and the system $P$, we say that
$P$ implements $H$ (written $P \propto H$) if
\begin{itemize} 
\item
$P$ is a correct orchestration composition and
\item
given a sequence $w$ of labels of the kind $a_{r \rightarrow s}$,
if $P \stackrel{w \surd}{\Longrightarrow} P'$
then there exists $H'$ such that $H \stackrel{w \surd}{\longrightarrow} H'$.
\end{itemize}
\end{definition}

Note that it is not necessary for an implementation
to include all possible conversations admitted
by a choreography.

\begin{example}{\bf (Implementation of Buyer/Seller/Bank)}
As an example, we present a possible implementation
of the choreography reported in the Example~\ref{BSB}.
\[
\begin{array}{l}
\ [\overline{\mathit{Request}}_{\mathit{Seller}} ; \mathit{Offer} ;
\overline{\mathit{Payment}}_{\mathit{Bank}} ; \mathit{Receipt}
]_{\mathit{Buyer}}\ 
\pa \\
\ [\mathit{Request} ; (\overline{\mathit{Offer}}_{\mathit{Buyer}} \parop                \overline{\mathit{PayDescr}}_{\mathit{Bank}});\mathit{Confirm}]_{\mathit{Seller}} 
\ \pa \\
\ [\mathit{PayDescr} ; \mathit{Payment}; (\overline{\mathit{Receipt}}_{\mathit{Buyer}} \parop  \overline{\mathit{Confirm}}_{\mathit{Seller}})]_{\mathit{Bank}}
\end{array}
\]

%Note that in this implementation we assume that the 
%travel agency always replies positively to the request
%of the client sending the $Confirmation$ message.
\end{example}

We now present the notion of choreography projection, which yields an orchestration $C$ for each role of a choreography $H$.
The definition is very simple thanks to the idea, we introduced in \cite{SC07}, of projecting communication atoms and then applying homomorphism over all the algebra operators.

\begin{definition}{\bf (Choreography projection)} 
\label{def:proj}
Given a choreography $H$, the projection 
$H$ on the role $r$, denoted with $\sem{H}_r$, is defined 
inductively on the syntax of $H$ in such a way that
\[
\sem{a_{r \rightarrow s}}_t\ =\
\left \{
\begin{array}{ll}
%\tau;
\overline{a}_s \qquad  & \mbox{if $t=r$} \\
a                           & \mbox{if $t=s$} \\ 
\pisucc                     & \mbox{otherwise}
\end{array}
\right .
\]
and that it is a homomorphism with respect to all
operators.
\end{definition}

It is interesting to observe that given a choreography $H$,
the system obtained composing its 
projections is not ensured to be an implementation of $H$.
For instance, consider the choreography
$
a_{r \rightarrow s}\ ;\ b_{t \rightarrow u} 
$.
The system obtained by projection is 
$
[\overline{a}_s]_r\ \pa\ [a]_s\ \pa\ [\overline{b}_u]_t\ \pa\ [b]_u
$.
Even if this is a correct composition of orchestrations,
it is not an implementation of $H$ because it comprises
the conversation $b_{t \rightarrow u} a_{r \rightarrow s}$ which
is not admitted by $H$.

The problem is not in the definition of the projection,
but in the fact that the above choreography
cannot be implemented preserving the 
message exchanges specified by the choreography. 
In fact, in order to guarantee that
the communication between $t$ and $u$ is executed after the
communication between $r$ and $s$, it is necessary to 
add a further message exchange (for instance between
$s$ and $r$) which is not considered in the choreography.
%This problem has been already investigated in~\cite{CHY07}
%where a notion of well formed choreography is introduced, 
%and it is proved that well formed choreographies
%admit a correct projection.\footnote{The projection defined
%in~\cite{CHY07} is more complex than ours as their
%choreography calculus comprises also an explicit notion
%of session.}

%Nevertheless, the notion of well formed choreography
%in~\cite{CHY07} is rather restrictive. In particular, after 
%the execution of a message sent from the role
%$v$ to the role $z$, the subsequent message in the conversation 
%should be mandatorily emitted by $z$.
%For instance, the choreography 
%$a_{r \rightarrow s} ; b_{r \rightarrow s}$
%does not satisfy this constraint even if
%the system
%$[\tau;\overline{a}_s;\tau;\overline{b}_s]_r\ \pa \ 
%[a;b]_s$ obtained by projection is a correct 
%implementation.

%To be less restrictive than~\cite{CHY07},
In order to have the guarantee that the system obtained
by projection is consistent with the initial choreography,
it is reasonable to consider a subset of {\em well formed}
choreographies.
The most general and intuitive notion of well formedness, we introduced in \cite{SC07},
can be obtained by
considering only all those
choreographies for which the system obtained by
projection is ensured to be a correct implementation.
\begin{definition}\label{wellFormed}{\bf (Well formed choreography)}
A choreography $H$, defined on the roles $r_1,\cdots,r_n$,
is {\em well formed} if 
$
[\, \sem{H}_{r_1}\, ]_{r_1}\ \pa\ \cdots\ \pa\ [\, \sem{H}_{r_n}\,]_{r_n}
\propto\ H
$
\end{definition}
It is worthwhile to note that well formedness is
decidable. In fact, given a choreography $H$,
it is sufficient to take the corresponding 
system $P$ obtained by 
projection, then consider $P$ and $H$ as finite 
state automata, and finally check whether the language 
of the first automaton is included in the language of the 
second one.
Note that the terms $P$ and $H$ can be seen as finite
state automata
thanks to the fact that their infinite behaviours
are defined using Kleene-star repetitions instead 
of general recursion.
%*********************************************************************************
%*********************************************************************************
%\subsubsection{Connected Choreographies}
%*********************************************************************************
%*********************************************************************************
This decidability result clearly follows from the fact that
we restrict to finite state choreographies. 

In the literature, 
syntactic versions of well formedness exist
(see e.g.~\cite{CHY07,sefm08,tgc08}). 
A sound charcterization of well-formedness is obtained, following~\cite{sefm08,tgc08}, by introducing a notion of connectedness
for our choreography calculus. The idea is to impose syntactic
restrictions in order to avoid the three possible ways in which
a system obtained by projection can have a different behaviour
w.r.t. its choreography: {\it connectedness for sequence}, related to the $H;H'$ operator and guaranteeing that 
an interaction of $H'$ cannot occur before
an interaction of $H$;
{\it unique point of choice}, related to the $H+H'$ operator and guaranteeing that 
all the roles involved in $H$ and $H'$ are aware of the selected branch; and
{\it no operation interference}, related to the $H|H'$ operator and guaranteeing that 
a message sent within
an interaction of one of $H$ or $H'$ cannot be
intercepted by a receive action in an interaction
of the other one (see~\cite{sefm08,tgc08} for details).

\begin{theorem}
Let $H$ be a choreography satisfying the %connectedness
{\it connectedness for sequence},
{\it unique point of choice} and
{\it no operation interference}
conditions.
Then $H$ is {\em well formed}.	
\end{theorem}

%\input{connectedcoreo}

%*********************************************************************************
\subsection{Behavioural Contracts and Contract Refinement}
%*********************************************************************************

Behavioural contracts make it possible to reason about protocol participant correct composition 
independently of the language (syntax and semantics) used for expressing the orchestration of a participant.
A behavioural contract is defined, basically, as a Labeled Transition System with the same set of labels we considered for the semantics of an orchestration (see, e.g.,~\cite{sfm09} for a precise definition).
An orchestration, therefore, gives rise to a behavioral contract as the labeled transition system obtained
by its semantics.

Behavioural contracts are important in the context of service oriented computing for dealing with the
problem of service retrieval.
Assuming that services expose in their interface an abstract description of their behaviour (a behavioural contract)
it is desirable to define an automatic procedure that can be used to check whether a service may correctly play a given 
role of a choreography.
It is therefore crucial to define an, as coarse as possible, notion of {\it contract refinement} that makes it possible
to establish if a discovered service can play a given role, based on the behavioural contract of the service and the
behavioural contract derived, by projection, from the choreography.
We, thus, define contract refinement as the coarsest preorder over behavioural contracts which preserves 
correct composition when applied to members of a set of behavioural contracts (one for each role of a choreography).
See \cite{fsen07,coord07,SC07,wsfm08,fundaInfo08,sfm09,mscs09,libroCubo} for formal definitions considering several form of communications/notions of correct composition.

%*********************************************************************************
%*********************************************************************************
%\subsubsection{Contract-based Service Discovery}
%*********************************************************************************
%*********************************************************************************

%\input{contracts}

%*********************************************************************************
\subsection{Introducing Distributed Dynamic Updates}
%*********************************************************************************

%\input{dynamicupdate}

% MACROS

%\newcommand{\Rule}[2]{\frac{#1}{#2}}
\newcommand{\Rule}[2]{\displaystyle\frac{#1}{#2}}

\newcommand{\arr}[1]{\xrightarrow{#1}}

\newcommand{\lts}[1]{\arr{#1}}

\newcommand{\updIndex}[3]{{#1} \{ {#2}:\;{#3}\}}
\newcommand{\upd}[3]{\updIndex{#1}{#2}{#3}}

\newcommand{\coupd}[3]{{#1}_{#2}\{#3\}}

\newcommand{\Did}[1]{(\textsc{#1})}

\newcommand{\pp}{\;\mathbf{|}\;}

\newcommand{\ppn}{\;\mathbf{|\!|}\;}

\newcommand{\pps}{\mathbf{|\!|}}

\newcommand{\zero}{\mathbf{0}}

\newcommand{\one}{\mathbf{1}}

\newcommand{\proj}[2]{\sem{#1}_{#2}}

\newcommand{\ol}[1]{\overline{#1}}
\newcommand{\m}[1]{\mathsf{#1}}

\newcommand{\outP}[4]{#1:\ol{#2}_{#3}\langle #4\rangle}
\newcommand{\inP}[4]{#1:{#2}_{#3}(#4)}

Dynamic updates are specified, similarly as in Section~\ref{s:calculi}, by defining \emph{scopes} $X[H]$ (where $X$ is a scope name) that delimit  
parts of choreographies that, at runtime, may be replaced by a new 
choreography, coming from  either inside or outside the system. 
Updates coming from outside may
be decided by the user through some adaptation interface, by some
manager module, or by the environment. In contrast, updates coming from inside
represent self-adaptations, decided by a part of the system towards itself
or towards another part of the system, usually as a result of some
unsatisfactory interaction. Updates from outside and
from inside are indeed quite similar, e.g., an update decided by a
manager module may be from inside if the manager behavior is part of
the choreography term, from outside if it is not.

Scopes $X[H]$ are updated by means of $\update{X}{H'}$  operators similarly as described in Section~\ref{s:calculi}.
However, for the sake of simplicity, here we just consider updates $\update{X}{H'}$ where $H'$ does not have holes $\bullet$, i.e. 
in Section~\ref{s:calculi} this corresponds to the case in which an update pattern $U$ is just a term $P$ (without any $\bullet$).
Moreover, we assume $\update{X}{H'}$ to just update the content of scope $X[H]$, thus leading to $X[H']$. 
Therefore, in terms of the machinery in Section~\ref{s:calculi}, this corresponds to having scopes $\component{a}{Q}$, for some $Q$, to be updated by $\update{a}{U}$ where $U$ is $\component{a}{P}$ for some $P$, thus $\component{a}{Q}$ is replaced by $\fillcon{\component{a}{P}}{Q} = \component{a}{P}$.
According to the classification introduced in Section~\ref{s:calculi}, updates of this kind fall in the category of {\em Unguarded} update patterns  (those of \evold{2} language) but not in that of {\em Preserving} update patterns (those of \evold{3} language). This because in the latter category update patterns are required to have one hole $\bullet$.
Moreover, the language topology is not static in that, when a named scope $X[H]$ is updated by means of $\update{X}{H'}$, $H'$ can have a different set of named scopes with respect to $H$ and, moreover, due to usage of the ``$;$'' operator, named scopes can be created and destroyed at run-time. 

Differently from process calculi considered in Section \ref{s:calculi} (\evol{} family), here scope updates are not applied by channel based communication (that would not make sense in the context of a choreography, where communication is just a basic action $a_{r \rightarrow s}$), but by replacement. Thus $\update{X}{H'}$ updates all scopes $X[H]$ (with the same name $X$) occurring anywhere in the choreography. This also justifies the change of notation from $\component{a}{P}$/$\update{a}{U}$ to $X[H]$/$\update{X}{H'}$: in the former $a$ represents a channel on which to communicate, in the latter $X$ represents a syntactic variable to replace.
Moreover, notice that, now when a scope $X[H]$ is updated, we are actually performing a {\it distributed update} that encompasses all the roles involved in $H$: at the orchestration level, as we will see, we have an $X$ scope for each of such roles and an update on $X$ must update all of them in a distributed way.

%%% * e ; a livello di linguaggio anziche ``.'' e bang e parallelo sincronizzante!
%%% meglio discutere di questo in fondo considerando anche update distribuito!

Formally speaking, the Choreography Calculus 
%The idea 
is extended %the calculus 
with {\em scopes} and
{\em updates} as follows (with respect to the notation above we add roles and we omit $\tilde{\;\;}$ in updates):

\begin{displaymath}
  \begin{array} {rllllllllll}
    H ::= & ... \\
    \quad | \quad & X:T[H]	 &\qquad (\text{scope}) &
    \quad | \quad & \upd{X}{r}{H}	 &\qquad (\text{update})
  \end{array}
\end{displaymath} 

\noindent
where $X$ is used to range over a set of {\em scope names} 
and $T$ is a set of roles. 
Construct $X:T[H]$ defines a scope named $X$
 currently executing choreography $H$ --- the name is needed to designate it as a target for a particular
adaptation. Type $T$ is the set
of roles (possibly) occurring in the scope. This is needed since a
given update can be applied to a scope only if it specifies how all
the involved roles are adapted. Operator $\upd{X}{r}{H}$ defines
\emph{internal updates}, i.e., updates offered by a participant of the
choreography. Here $r$ denotes the role offering the
update, $X$ is the name of the target scope, and $H$ is the new
choreography.

The operational semantics for the new Choreography Calculus with updates
is defined only for a proper subset of well defined choreographies.\footnote{We refer
the reader to~\cite{beat} for a formal definition of this subset of well
defined choreographies, here we simply informally report the imposed limitations.}
%
%Not all choreography terms generated by the syntax above are useful
%choreographies.  To formally define the choreography terms which
%actually represent choreographies, we rely on some auxiliary
%definitions.  The set of roles inside a choreography term $H$, denoted
%$roles(H)$, is defined inductively as follows:
%\begin{eqnarray*}
%roles(\chor a{r_1}{r_2}) & = & \{r_1, r_2\} \qquad\qquad ~\, roles(\upd{X}{r}{H})  =  \{r\}  \\
%roles(X:T[H]) & = & T \cup roles(H) \qquad
%roles(H^*)  =  roles(H) \\
%roles(H_1\; ;\; H_2) & = &roles(H_1\ \pp \ H_2) = roles(H_1 + H_2) = roles(H_1) \cup roles(H_2) \\
%roles(\one) & = & roles(\nil) = \emptyset
%\end{eqnarray*}
%
%This 
%includes
%roles of actions, i.e., $r_1$, $r_2$ in ${\chor a{r_1}{r_2}}$, 
%types of
%scopes, i.e., $T$ in $X:T[C']$, and
%roles originating update prefixes, i.e., 
%Notice that for 
%$\upd{X}{r}{H}$ we consider role $r$ but not the roles in $H$.
%This is because the choreography $H$ could be injected,
%and then executed, in a different part of the choreography.
% may be
%%correspond to an 
%externally updated with some different choreography term.
%Further, we shall say that a choreography term $C$ is \emph{well-typed} 
%if:
%\begin{enumerate}[(i)]
%\item for every scope $X:T[C']$ occurring in $C$ 
%we have $roles(C') \subseteq T$ and
%\item every update prefix $\upd{X}{r}{C''}$ occurring in $C$ is such that $roles(C'') \subseteq T$.
%\end{enumerate}
%
%Not all choregraphies are correct; in particular, 
First of all, type $T$ related conditions are added in order to guarantee that:
every scope $X:T[H']$ occurring in $H$ has the same type $T$, denoted with $\mathit{type}(X)$; and
every update $\upd{X}{r}{H}$ injects a choreography $H$ that considers only
roles explicitly indicated in the type $T$ of the updated
part(s) $X:T[H']$ (at least one scope $X$ is required to occur in the choreography). Moreover, also syntactic restrictions are imposed
in order to guarantee that it will never occur that two distinct
scopes with the same name are contemporaneously active.
This is necessary for the following reason: when a scope is projected,
it is distributed among several independent roles running in parallel, 
and in order to avoid interferences between two distinct scopes with the
same name $X$ we assume that only one scope $X$ can be active
at a time.

The operational semantics %for choreographies with updates 
is defined
by adding the rules dealing with {\em scopes} and {\em updates}
reported in Table~\ref{table:semChoreo}. The transition labels $\eta$
now include the label $\upd{X}{r}{H}$ indicating the execution
of an update action. 
%We now define the semantics of choreography terms via a labeled
%transition system.  As in the syntax, the most interesting part of the
%semantics concerns update constructs.  Recall that $T$ is a set of
%roles.  
In the rules, we use $H [H'/X]$ to denote the
substitution that replaces all scopes $X:T[H'']$ with name $X$
occurring in $H$ (not inside update prefixes) with $X:T[H']$.  

\begin{table}[t]
\begin{displaymath}
\begin{array}{c}
%    \Did{One} ~ \Rule {} { \one \lts{\surd} \nil} \qquad 
%    \Did{Comm} ~ \Rule {} { \chor a{r_1}{r_2} \lts{\chor {a^{\varepsilon}}{r_1}{r_2}} \one} \qquad 
%    \Did{Seq} ~  \Rule {C_1\lts{\chor {a^{\omega}}{r_1}{r_2}} C_1'  }
%    {C_1;\ C_2 \lts{\chor {a^{\omega}}{r_1}{r_2}} C_1';C_2}\\ [6mm]
%    \Did{SeqTick} ~ \Rule {C_1 \lts{\surd} C_1' \qquad C_2 \lts{\alpha} C_2' }
%    {C_1;\ C_2 \lts{\alpha} C_2'}
%    \qquad
%    \Did{Par} ~
%    \Rule {C_1\lts{\chor {a^{\omega}}{r_1}{r_2}} C_1' }
%    {C_1\ \pp\ C_2  \lts{\chor {a^{\omega}}{r_1}{r_2}} C_1'\ \pp\ C_2}
%    \\[6mm]
%    \Did{ParTick}~
%    \Rule {C_1\lts{\surd} C_1'  \qquad C_2\lts{\surd} C_2' }
%    {C_1\ \pp\ C_2  \lts{\surd} C_1'\ \pp\ C_2'}
%    \qquad
%    \Did{Cho} ~
%    \Rule {C_1\lts{\alpha} C_1' }
%    {C_1 +  C_2  \lts{\alpha} C_1'}
%    \\[6mm]
%    \Did{Star} ~
%    \Rule {C\lts{\chor {a^{\omega}}{r_1}{r_2}} C' }
%      {C^* \lts{\chor {a^{\omega}}{r_1}{r_2}} C';\ C^*}
%    \qquad
%    \Did{StarTick} ~ \Rule {} { C^* \lts{\surd} \nil}  
   \\[6mm]
   \Did{CommUpd} ~ \Rule {} { \upd{X}{r}{H} \lts{\upd{X}{r}{H}} \one}
    \qquad
    \Did{SeqUpd} ~ 
    \Rule {H_1\lts{\upd{X}{r}{H}} H_1'} %\qquad C_2\lts{\coupd{X}{r}{C}} C_2'} 
    {H_1;\ H_2 \lts{\upd{X}{r}{H}} H_1';(H_2 [H/X])}
    \\[6mm]
    \Did{ParUpd} ~ 
    \Rule {H_1\lts{\upd{X}{r}{H}} H_1'} %\qquad C_2\lts{\coupd{X}{r}{C}} C_2'} 
    {H_1\ \pp\ H_2 \lts{\upd{X}{r}{H}} H_1'\ \pp\ (H_2 [H/X]) }
    \quad
    \Did{StarUpd} ~
    \Rule {H_1\lts{\upd{X}{r}{H}} H_1'} %\qquad C\lts{\coupd{X}{r}{C}} C''} 
    {H_1^* \lts{\upd{X}{r}{H}} H_1';\ (H_1 [H/X])^*}
    \\[6mm]
    \Did{ScopeUpd} ~
    \Rule {H_1\lts{\upd{X}{r}{H}} H_1'}  
    {X:T[H_1] \lts{\upd{X}{r}{H}} X:T[H] }
    \quad
    \Did{ScopeComm} ~
    \Rule {H_1\lts{\chor {a}{r_1}{r_2}} H_1'}  
    {X:T[H_1] \lts{\chor {a}{r_1}{r_2}} X:T[H_1'] }
    \\[6mm]
    \Did{Scope} ~ 
    \Rule {H_1\lts{\eta} H_1' }
    {X:T[H_1] \lts{\eta} X:T[H_1'] } \;\; 
\eta \neq \upd{X}{r}{H} \; \mbox{for any} \; r,H 
%\wedge 
%\alpha \neq \chor {a^{\omega}}{r_1}{r_2}  \; \mbox{for any} \; w,r_1,r_2

\end{array}
\end{displaymath}
\caption{Semantics of Choreographies with updates}\label{table:semChoreo}
\end{table}

%\begin{definition}\label{semUpdChoreo}
%The semantics of choreography terms is the smallest labeled transition
%system closed under the rules in Table~\ref{} and Table~\ref{table:semChoreo}
%(where Rule~\Did{Comm} in Table~\ref{table:semChoreo} replaces
%the corresponding one in Table~\ref{}) .
%\end{definition}

We briefly comment 
%on the rules in Table~\ref{table:semChoreo}.
%Rules in the first four rows of the table are standard (cf.~\cite{contracts}). 
%Rule~\Did{One} defines termination for the empty choreography term.
%Rule~\Did{Comm} executes an
%interaction, making it visible in the label. While rule~\Did{Seq} allows the
%first component of a sequential composition to compute, rule~\Did{SeqTick} allows it to terminate, starting the execution
%of the second component. Rule~\Did{Par} allows parallel components to
%interleave their executions. Rule \Did{ParTick} allows parallel
%components to synchronize their termination. Rule \Did{Cho} selects a
%branch in a nondeterministic choice. Rule \Did{Star} unfolds the
%Kleene star. Note that the unfolding may break uniqueness of scopes
%with a given name---we will come back to this point later on.  
%Rule~\Did{StarTick} defines termination of a Kleene star.
%
the %The remaining 
rules in Table~\ref{table:semChoreo}.
% that deal with adaptation. 
Rule \Did{CommUpd} makes
 an internal update available, moving the information to the
label. Updates propagate through sequence, parallel composition,
and Kleene star using rules \Did{SeqUpd}, \Did{ParUpd}, and
\Did{StarUpd}, respectively. Note that, while propagating, the update is
applied to the continuation of the sequential composition, to parallel
terms, and to the body of Kleene star. Notably, the update is
applied to both enabled and non enabled
occurrences of the desired scope. Rule~\Did{ScopeUpd} allows a scope
to update itself (provided that the names coincide), while propagating
the update to the rest of the choreography. %Finally, r
Rule~\Did{Scope} allows a scope to compute.

\begin{example}\label{AdaptableBSB}
{\bf (Adaptable Buyer/Seller/Bank)}
Here, we consider a version of the Buyer/Seller/Bank example
discussed in the Example~\ref{BSB} where it is possible to update
the payment interaction between the buyer and the bank by using, 
for instance, a new version %$PaymentVISA$ 
of the payment protocol 
according to which the buyer sends its VISA code
to the bank and the bank subsequently confirms its correctness. 
Let us consider the following choreography composed of three
roles: {\em Buyer}, {\em Seller} and {\em Bank}
\[
\begin{array}{l}
\mathit{Request}_{\mathit{Buyer} \rightarrow \mathit{Seller}};
(\ \mathit{Offer}_{\mathit{Seller} \rightarrow \mathit{Buyer}} \parop
\mathit{PayDescr}_{\mathit{Seller} \rightarrow \mathit{Bank}}\ ); \\
\hspace{.5cm} X\{\mathit{Buyer},\mathit{Bank}\}[\mathit{Payment}_{\mathit{Buyer} \rightarrow \mathit{Bank}}];
(\ \mathit{Confirm}_{\mathit{Bank} \rightarrow \mathit{Seller}} \parop 
\mathit{Receipt}_{\mathit{Bank} \rightarrow \mathit{Buyer}}\ )
\end{array}
\]
According to the operational semantics defined above,
this choreography could, for instance, perform the initial $\mathit{Request}$
interaction and then receives an external
update:
$$\upd{X}{r}{VISAcode_{Buyer \rightarrow Bank};VISAok_{Bank \rightarrow Buyer}}$$ 
and then becomes the following choreography:
\[
\begin{array}{l}
(\ \mathit{Offer}_{\mathit{Seller} \rightarrow \mathit{Buyer}} \parop
\mathit{PayDescr}_{\mathit{Seller} \rightarrow \mathit{Bank}}\ ); \\
\hspace{.5cm} X\{\mathit{Buyer},\mathit{Bank}\}[\mathit{VISAcode}_{\mathit{Buyer} \rightarrow \mathit{Bank}};\mathit{VISAok}_{\mathit{Bank} \rightarrow \mathit{Buyer}}];
(\ \mathit{Confirm}_{\mathit{Bank} \rightarrow \mathit{Seller}} \parop 
\mathit{Receipt}_{\mathit{Bank} \rightarrow \mathit{Buyer}}\ )
\end{array}
\]
\end{example}

%We now move to the analysis of a corresponding Orchestration Calculus
%with updates. 

We are now ready to present the definition of our 
Orchestration Calculus extended with operators for dynamic updates:
%Since choreographies are at the very high level of abstraction,
%defining a description of the same system nearer to an actual
%implementation is of interest. In particular, for each participant in
%a choreography (also called \emph{endpoint}) we would like to describe
%the actions it has to take in order to follow the choreography.  The
%syntax of endpoint processes is as follows:
%\subsubsection{Syntax}
\begin{displaymath}
  \begin{array} {rllllllllll}
    C ::= & ...\\
    \quad | \quad & X[C]^F            &\qquad (\text{scope}) & 
    \quad | \quad & \upd{X}{(r_1,\ldots,r_n)}{C_1,\ldots,C_n}      &\qquad (\text{update})
%    \quad | \quad & \langle P\rangle  &\qquad (protection)\\
% nil??
  \end{array}
\end{displaymath} 
where $F$ is a {\it flag} that is either $A$, standing for active (running) scope, or $\varepsilon$, denoting a scope still to be started ($\varepsilon$ is omitted in the following).
%
%As for choreographies, endpoint processes contain some standard operators and
%some operators dealing with adaptation. Communication is performed by
%$\overline a_r$, denoting an output on channel $a$ towards
%participant $r$.
%Dually,  $a_r$ denotes an input
%from participant $r$ on channel $a$. 
%Intuitively, an output $\overline a_r$ in role $s$ and an input $a_s$ in role $r$ should synchronize.
%%even when $r$ and $s$ denote different roles.
%%Communication actions can be composed in sequence ;, in parallel $\pp$, and using nondeterministic choice $+$. 
%Two endpoint processes $P_1$ and $P_2$ can be composed in sequence ($P_1 \ ; \ P_2$), in parallel ($P_1 \ \pp \ P_2$), and using nondeterministic choice ($P_1 + P_2$).
%Endpoint processes can be iterated using a Kleene star $\mbox{}^*$.
%The empty endpoint process is denoted by $\one$ and the deadlocked endpoint process
%%(needed only for the definition of the semantics) 
%is denoted by $\nil$.
%
%Updates are applied to scopes. 
$X[C]^F$ denotes a scope named $X$
executing $C$. $F$ is a flag distinguishing scopes whose
execution has already begun ($A$) from scopes which have
not started yet ($\varepsilon$). In order for scopes to become active,
orchestrations starting execution of scopes with the same name $X$ must synchronize.
Also, when all participants in a scope $X$ complete their 
respective executions, a synchronisation is needed in order
to synchronously remove the scope. 
The update operator
$\upd{X}{(r_1,\ldots,r_n)}{C_1,\ldots,C_n}$ provides an update for scope named $X$, involving roles
$r_1,\ldots,r_n$. The new behaviour for role $r_i$ is $C_i$.

As in the previous sections, systems are of the form $[C]_r$, where $r$ is the name of
the role and $C$ its behaviour. Systems, denoted $P$, are 
obtained by parallel
composition of 
orchestrations: 
%Endpoints can be composed into systems using parallel
%composition $\pps$. Thus the syntax of endpoints is as follows.
\begin{displaymath}
  \begin{array} {rllllllllll}
    P\ ::= & [ C ]_r               %&\qquad (\text{endpoint}) & 
    \quad |  & P \pps P     %&\qquad (\text{parallel system})
\\
  \end{array}
\end{displaymath} 

%As for choreographies, not all systems are endpoint specifications.
%By a slight abuse of notation we extend $type(X)$ to 
%endpoints associating a set of roles to each scope name $X$.
%Endpoint specifications are defined as follows.
%
%\begin{definition}
%A system $S$ is an endpoint specification if the following conditions hold: 
%\begin{enumerate}[(i)]
%\item  no active scopes are present
%\item endpoint names are 
%unique
%it 
%does not include multiple roles with the same name $r$ (we denote with $roles(S)$ the set of role 
%names of the roles included in $S$), 
%\item all roles $r$ occurring in terms of the form $\overline a_r$, $a_r$, or such that $r \in 
%type(X)$ for some scope $X$ are endpoints of $S$
%%scopes with name $X$ con occur in $S$ only if $type(X) \subseteq roles(S)$, outputs 
%%$\overline a_r$ and inputs $a_r$ included in $S$ are such that $r \in roles(S)$,
%\item a scope with name $X$ con occur (outside updates) only in endpoints $r \in 
%type(X)$
%\item every update has the form $\upd{X}{type(X)}{P_1,\ldots,P_n} $ 
%%we have $\{ 
%%r_1,\ldots,r_n\} = type(X)$, % \subseteq type(X)$,
%\item outputs $\overline a_r$ and inputs $a_r$ included in $\upd{X}{type(X)}{P_1,\ldots,P_n}$ are 
%such that $r \in type(X)$.
%%oppure $\in \{ r_1,\ldots,r_n\}$???
%%a scope with name $X$ con occur in $P_i$ associated to role $r_i$ not inside update prefixes only 
%% if $r_i \in type(X)$. 
%\end{enumerate}
%\end{definition}

In this presentation, we do not formally define a semantics for
orchestrations: we just point out that it should include labels
corresponding to all the labels of the semantics of choreographies, 
plus some additional labels corresponding to partial
activities, such as an input. We also highlight the fact that all
scopes that correspond to the same choreography scope evolve
together: they are endowed with {\it scope start} transitions (transforming a scope from
inactive to active, setting its flag $F$ to $A$) that are synchronized; and with {\it scope end}
transitions (syntactically removing the entire scope) that are synchronized as well.  The fact that choreographies feature at
most one scope with a given name is instrumental in ensuring this
property.

%\paragraph{Projection.} Since choreographies provide system
%descriptions at an high level of abstraction and endpoint
%specifications provide more low level descriptions, a main issue is to
%derive from a given choreography an endpoint specification executing
%it. This is done using the notion of \emph{projection}.

We now discuss how to extend the notion of projection presented
in Definition~\ref{def:proj} for the case without updates.

\begin{definition}{\bf (Projection for choreographies with updates)}
The projection of a choreography $H$ on a role $r$, denoted by $\proj{H}{r}$, is defined 
as in Definition~\ref{def:proj} plus the clauses below for scopes and updates:
%\begin{itemize}
%\item 
%$\proj{\chor a{r_1}{r_2}}{r} = {\overline a_{r_2}}$, if $r = r_1$;\\
%$\proj{\chor a{r_1}{r_2}}{r} = a_{r_1}$ if $r = r_2$;\\ $\proj{\chor a{r_1}{r_2}}{r} = \one$, otherwise.
%\item
%$\proj{\upd{X}{r'}{C}}{r} = \upd{X}{(r_1,\ldots,r_n)}{\proj{C}{r_1},\ldots,\proj{C}{r_n}}$, where $\{r_1,\ldots,r_n\}= type(X)$, if $r = r'$;\\
%$\proj{\upd{X}{r'}{C}}{r} = \one$, otherwise.
%\item
%$\proj{X:T[C]}{r} = X[\proj{C}{r}]$ if $r \in type(X)$;\\
%$\proj{X:T[C]}{r} = \one$, otherwise.
%\end{itemize}
\begin{eqnarray*}
%\proj{\chor a{r_1}{r_2}}{r} & = & \begin{cases}{\overline a_{r_2}} & \text{if $r = r_1$}\\
%												  a_{r_1} & \text{if $r = r_2$}\\ 
%												  \one &  \text{otherwise}\end{cases} \\
\proj{\upd{X}{r'}{H}}{r} & = & \begin{cases}\upd{X}{(r_1,\ldots,r_n)}{\proj{H}{r_1},\ldots,\proj{H}{r_n}}  
\text{with $\{r_1,\ldots,r_n\}= \mathit{type}(X)$} &
\!\!\!\!\text{if $r = r'$}\\
\one & 
\!\!\!\!\!\text{otherwise}\end{cases} \\
\proj{X:T[H]}{r} &= & \begin{cases} X[\proj{H}{r}] & \text{if $r \in \mathit{type}(X)$}\\
\one &  \text{otherwise} \end{cases}
\end{eqnarray*}
%
%
%and is an homomorphism on the other operators.
%The endpoint specification resulting from a choreography $C$ is
%obtained by composing in parallel roles $[\![\proj{C}{r}]\!]_r$, where $r
%\in roles(C)$. 
\end{definition}

\begin{example}
%\label{AdaptableBSB}
%{\bf (Adaptable Buyer/Seller/Bank)}
We now present the projection of the choreography 
in the Example~\ref{AdaptableBSB} (we omit unnecessary $\one$ terms):
\[
\begin{array}{l}
\ [\overline{\mathit{Request}}_{\mathit{Seller}} ; \mathit{Offer} ;
X[\overline{\mathit{Payment}}_{\mathit{Bank}}] ; \mathit{Receipt}
]_{\mathit{Buyer}}\ 
\pa \\
\ [\mathit{Request} ; (\overline{\mathit{Offer}}_{\mathit{Buyer}} \parop                \overline{\mathit{PayDescr}}_{\mathit{Bank}});\mathit{Confirm}]_{\mathit{Seller}} 
\ \pa \\
\ [\mathit{PayDescr} ; X[\mathit{Payment}]; (\overline{\mathit{Receipt}}_{\mathit{Buyer}} \parop  \overline{\mathit{Confirm}}_{\mathit{Seller}})]_{\mathit{Bank}}
\end{array}
\]
It is interesting to note that the projection clearly identifies where
a possible update of the payment should have an effect; namely,
only the roles $\mathit{Buyer}$ and $\mathit{Bank}$ are affected by the update
in precise parts of their behaviour.
For instance, if 
$\upd{X}{(\mathit{Buyer},\mathit{Bank})}{\big(\overline{\mathit{VISAcode}}_{\mathit{Bank}};\mathit{VISAok}\big),\big(\mathit{VISAcode};\overline{\mathit{VISAok}}_{\mathit{Buyer}}\big)}$ 
is executed after the first $\mathit{Request}$
interaction occurs, then the system becomes:
\[
\begin{array}{l}
\ [\mathit{Offer} ;
X[\overline{\mathit{VISAcode}}_{\mathit{Bank}};\mathit{VISAok}] ; \mathit{Receipt}
]_{\mathit{Buyer}}\ 
\pa \\
\ [\mathit{(\overline{Offer}}_{\mathit{Buyer}} \parop                
\overline{\mathit{PayDescr}}_{\mathit{Bank}});\mathit{Confirm}]_{\mathit{Seller}} 
\ \pa \\
\ [\mathit{PayDescr} ; X[\mathit{VISAcode};\overline{\mathit{VISAok}}_{\mathit{Buyer}}]; (\overline{\mathit{Receipt}}_{\mathit{Buyer}} \parop  \overline{\mathit{Confirm}}_{\mathit{Seller}})]_{\mathit{Bank}}
\end{array}
\]
where the projections of the new protocol are precisely injected
in the behaviour of the affected roles.
\end{example}

%As an example, the endpoint projection obtained from the prescribe
%choreography introduced in \S\ref{sec:chor} is $[\![ P_{\texttt{N}}
%]\!]_N \pps [\![ P_{\texttt{D}} ]\!]_D \pps [\![ P_{\texttt{P}}
%]\!]_P$
%%
%where processes $P_{\texttt{N}}$, $P_{\texttt{D}}$, and $P_{\texttt{P}}$ are as follows (we omit 
%unnecessary $\one$ processes):
%\begin{eqnarray*}
%  P_{\texttt{N}} & = & X[((prescribe_D)^+ \; ; \; ( sign_D  + up_D \;) ; \; \overline{trust}_D)^+] \; 
%  ; 
%  \;\overline{medicine}_P\\
%  P_{\texttt{D}} & = & X[((\overline{prescribe}_N)^+ \; ; \; (\overline{sign}_N + 
%  X_{D,N}\{\overline{sign}_N, 
%  sign_D\} \; ; \; \overline{up}_N );trust_N)^+] \\
%  P_{\texttt{P}} & = & medicine_N
%\end{eqnarray*}

%One can see that the system obtained by projection
%is an endpoint specification. 
Ideally, traces of the projected system
should correspond to the traces of the original
choreography. Actually, we conjecture that this occurs for
choreographies satisfying connectedness conditions 
obtained by extending those already discussed in Section~\ref{proj}.
We finally point out two main aspects of the expected correspondence result
between choreographies and their projections in the case of the calculi
extended with dynamic updates.  First, labels
$\upd{X}{r}{H}$ %and $\coupd{X}{r}{C}$
of transitions of the choreography should be mapped to labels
%$[\upd{X}{(r_1,\ldots,r_n)}{H_1,\ldots,H_n}]_r$ %and $[\coupd{X}{(r_1,\ldots,r_n)}{P_1,\ldots,P_n}]_r$, respectively,
of the transitions of the Orchestration Calculus
obtained by appropriate label projections.
% %(see the Appendix)
%of the endpoint specification obtained by projection,
%where $type(X) = \{r_1,\ldots,r_n\}$ and $H_1 = \proj{C}{r_1},\ldots,
%H_n = \proj{H}{r_n}$ are obtained %, themselves,
%by projection from $H$. 
Second, orchestration traces should not consider
unmatched input and output labels.

\section{Related Work}\label{s:related}
%*********************************************************************************
%*********************************************************************************

The $\evol{}$ calculus
is related to \emph{higher-order} process calculi such as, e.g.,
 the higher-order $\pi$-calculus \cite{San923},  Kell \cite{SchmittS04}, and Homer \cite{Mikkel04}.
 (Further comparisons between  $\evol{}$ and other calculi and languages can be found in~\cite{lmcs}.)
In such calculi, % In a higher-order setting,  %process calculi,  
  processes can be passed around, and so 
communication involves term instantiation, as in the $\lambda$-calculus.
Update actions  in \evol{}  are a  form of term instantiation: 
%as we elaborate in \cite{BGPZFACS}, 
they can be seen as a streamlined version of  
 the \emph{passivation} operator of Kell and Homer, which allows 
to suspend a running process. 
It would be interesting to investigate if the results and techniques 
developed in this paper can apply to Kell and Homer (or to some interesting fragments of them).

Concerning theories related to choreographies and orchestrations/contracts, among our main contributions we can mention:
(i) The formalisation of the relationship between 
global choreographic descriptions and systems obtained 
as parallel compositions of peers,
(ii) The definition of suitable notions of behavioural contract refinement, and 
(iii) The proposal of mechanisms for dynamic 
updates for both the choreography and the orchestration calculi.
Concerning (i), we have defined
well-formedness for choreographies based on 
natural projection and the notion of implementation.
We have introduced the simple technique of obtaining orchestrations
by projection, defining it for communication actions and then 
exploiting homomorphism over all the process algebraic operators.
Moreover our approach leads to a more general notion of well-formedness
w.r.t. other approaches like, e.g.,~\cite{CHY07},
where it is defined in terms of three
connectedness constraints similar to those we
have mentioned in Section~\ref{proj}.
Concerning (ii), our main contribution is related to 
the idea of refining all peers guaranteeing that
all of them continue to reach local success. 
This differs from other popular approaches,
like those initiated by Fournet et al.~\cite{FHRR04} 
or Padovani et al.~\cite{CCLP06}, where the focus is 
on the success of one specific peer (usually,
the so-called, {\em client}).
Concerning (iii), it is worth to mention
that~\cite{ivan} has been a source of inspiration
for the present work:
the main difference is 
our choice of expressing adaptation in terms of scopes 
and code update constructs, rather than using rules.
This approach appears more adequate for the definition 
of a general theory of behavioural typing to be used 
on more general languages where 
multiple protocols/choreographies 
can interleave inside the same program.

%*********************************************************************************
%*********************************************************************************
\section{Conclusion}\label{s:conclusion}
%*********************************************************************************
%*********************************************************************************

We just present some observations about the reported results and remarks concerning current/future work. 

Choreography and orchestration languages of Section \ref{chorupd} make use, as common in this context,
of Kleene-star instead of general recursion (or replication as for $\evol{}$ calculi considered in Section \ref{s:calculi}).
As a consequence (see Section \ref{proj}), they always give rise to finite-state transition systems,
for which verification problems are decidable. 
%as usual for choreography and orchestration languages, considered in Section \ref{chorupd} are finite-state.
Given that update mechanisms we introduced in this context belong to the class of those considered in the
$\evold{2}$ language of Section \ref{s:calculi}, verification of properties like bounded adaptability or formulae of a restricted 
logic like $\Fres$ could be still decidable even if we extend such languages with some more expressive form of recursion.
This, for instance, could introduce the possibility, similarly as for the $\evol{}$ family, to generate new participants (or participant instances) at run-time.
Notice that such a correspondence between decidability results for the $\evol{}$ family and for choreography and orchestration languages would be possible because, as common in the context of latter languages, name binders (CCS restriction) are not considered, which would otherwise make them Turing complete.

%Now some remarks concerning future work.
We are currently working on applying the theory of updatable choreographies/orchestrations in the context of session types for typing a concrete language with session spawning, where choreographies play the role of global types attached to sessions and we use orchestrations for checking, via typing rules, that the code actually conforms with the specified global types.
In this context, extending  our contract refinement theory \cite{fsen07,SC07,fundaInfo08,sfm09,libroCubo} to updatable choreographies/orchestrations would make it possible to define a notion of semantic subtyping.

\newcommand{\etalchar}[1]{$^{#1}$}

\providecommand{\urlalt}[2]{\href{#1}{#2}}
\providecommand{\doi}[1]{doi:\urlalt{http://dx.doi.org/#1}{#1}}

%\bibliographystyle{eptcs}
%\bibliography{bibliography}

\end{document}